\documentclass[twocolumn,showpacs,amsmath,amssymb,aps,prd,floats,nofootinbib]{revtex4-1}
\usepackage{bm}% bold math
\usepackage{graphicx}
\usepackage{times}
\usepackage{dcolumn}% Align table columns on decimal point
\usepackage{epsfig}
\usepackage{epstopdf}
\usepackage{color}

\newcommand{\EQ}{\begin{equation}}
\newcommand{\EN}{\end{equation}}
\newcommand{\EQA}{\begin{eqnarray}}
\newcommand{\ENA}{\end{eqnarray}}

\newcommand{\xx}{\mbox{\boldmath $x$} {}}
\newcommand{\nn}{\mbox{\boldmath $n$} {}}
\newcommand{\kk}{\mbox{\boldmath $k$} {}}
\newcommand{\KK}{\mbox{\boldmath $K$} {}}
\newcommand{\qq}{\mbox{\boldmath $q$} {}}
\newcommand{\bb}{\mbox{\boldmath $b$} {}}
\newcommand{\rr}{\mbox{\boldmath $r$} {}}
\newcommand{\llvec}{\mbox{\boldmath $\ell$} {}}
\newcommand{\mm}{\mbox{\boldmath $m$} {}}
\newcommand{\ttvec}{\mbox{\boldmath $t$} {}}

\newcommand{\ssvec}{\mbox{\boldmath $s$} {}}
\newcommand{\wwvec}{\mbox{\boldmath $w$} {}}

\newcommand{\kkk}{\hat{\bm{k}}}

\newcommand{\sss}{\hat{\bm{s}}}
\newcommand{\rrr}{\hat{\bm{r}}}
\newcommand{\lllvec}{\hat{\bm{\ell}}}
\newcommand{\wwhat}{\hat{\bm{\omega}}}

\begin{document}

\title{Primordial Magnetic Field Limits from CMB Trispectrum - Scalar Modes and Planck Constraints}
\author{Pranjal Trivedi}
\affiliation{
Department of Physics, Sri Venkateswara College, University of Delhi, Delhi 110021, India, \\ \& Department of Physics and Astrophysics, University of Delhi, Delhi 110007, India.}
\email{ptrivedi@physics.du.ac.in}
\author{Kandaswamy Subramanian}
\affiliation{IUCAA, Post Bag 4, Ganeshkhind, Pune 411 007, India.}
\email{kandu@iucaa.ernet.in}
\author{T. R. Seshadri}
\affiliation{Department of Physics and Astrophysics, University of Delhi,
Delhi 110007, India.}
\email{trs@physics.du.ac.in}

\date{\today}

\begin{abstract}
Cosmic magnetic fields are observed to be coherent on large scales and could have a primordial origin. 
Non-Gaussian signals in the cosmic microwave
background (CMB) are generated by primordial magnetic fields as the magnetic stresses and
temperature anisotropy they induce depend quadratically on the magnetic field. 
We compute 
the CMB scalar trispectrum 
on large angular scales, for nearly scale-invariant magnetic fields, sourced via the Sachs-Wolfe effect.
The trispectra induced by magnetic energy density and by magnetic scalar anisotropic stress are found to have
typical magnitudes of approximately 
$10^{-29}$ and 
$10^{-19}$, respectively. The scalar anisotropic stress 
trispectrum is also calculated in the flat-sky approximation and yields a similar result. 
Observational limits on CMB non-Gaussianity from the {\it Planck} mission data
allow us to set upper limits of $B_0 \lesssim 0.6 $ nG on the present value of the
primordial cosmic magnetic field. Considering the inflationary magnetic curvature mode 
in the trispectrum can further tighten the magnetic field upper limit to $B_0 \lesssim 0.05 $ nG.
These sub-nanoGauss constraints from the magnetic trispectrum are 
the most stringent limits so far on the strength of primordial magnetic fields, on megaparsec scales, 
significantly better than the limits obtained from the CMB bispectrum and the CMB power spectrum.

\end{abstract}

\maketitle

\section{Introduction}
\label{s:Intro}

Magnetic fields have been observed throughout the Universe, on all scales probed so far, 
from planets and stars to the large-scale magnetic fields detected in galaxies and galaxy clusters 
\cite{BeckWiel13,Beck09,Vogt05,Clarke01,DN13,BS05,KZ08,W02}. Both large-scale as well as stochastic components 
are present in magnetic fields observed in galaxies with magnitudes from a few to tens of microGauss.
Coherent magnetic fields of a similar strength are also observed in higher redshift galaxies \cite{Bernet08,Kronberg08}. 
In clusters of galaxies, 
stochastic magnetic fields of a few microGauss strength are present, correlated on ten kiloparsec scales 
\cite{Vogt05,Clarke01}.
Moreover, there is circumstantial evidence of an intergalactic magnetic field that is present 
over most of the cosmic volume,even in the voids of large scale structure. 
A lower bound of $10^{-16} - 10^{-15}$ Gauss for such a pervasive
intergalactic magnetic field has been derived from gamma-ray observations of blazars \cite{neronov10,Tavecchio10,Dolag11}. 

The origin as well as evolution of such large-scale magnetic fields remains an outstanding problem. 
Magnetic fields in collapsed structures can arise from dynamo amplification of seed magnetic fields \cite{BS05,KZ08,W02}.
The seed field could in turn be generated in astrophysical batteries \cite{SNC94,Kulsrud97,Gnedin2000,SethiGopal05}
or due to processes in the early universe 
\cite{TurnerWidrow88,Ratra92,MartinYokoyama08,Giovannini08,KS_AN10,Kandus_review11,
Vachaspati91,Banerjee_Jedamzik04,DiazGil08,Copi08,Kahniashvili10}.
Indeed, the recent gamma-ray observations suggesting a lower limit
to an all-pervasive intergalactic magnetic field 
\cite{neronov10,Tavecchio10,Dolag11}, would perhaps favour a primordial origin. 
A primordial magnetic field can be generated at inflation 
\cite{TurnerWidrow88,Ratra92,MartinYokoyama08,Giovannini08,DN13,KS_AN10,Kandus_review11}, 
or arise out of other phase transitions in the early Universe 
\cite{Vachaspati91,Banerjee_Jedamzik04,DiazGil08,Copi08,Kahniashvili10}. 
As yet there is no compelling mechanism which produces strong
coherent primordial fields.
Equally, the dynamo paradigm is not without its own challenges
in producing sufficiently coherent fields and sufficiently rapidly \cite{BS05,KZ08,W02}.
Therefore, it is useful to keep open the possibility that
primordial magnetic fields originating in the early universe play a
crucial role in explaining the observed cosmic magnetism.

In this context it is important to investigate every possible observable 
signature of the putative primordial magnetic field. Magnetic fields
give rise to scalar, vector and tensor metric perturbations as well as fluid
perturbations via the Lorentz force. 
Constraints on large scale primordial magnetic fields have already
been derived using the CMB temperature and polarization power spectra 
\cite{Planck_Parameters,Pao12,Pao10,Yam10,SL10,Giovannini_Kunze08,KS_AN06,Durrer_rev07} 
and Faraday rotation \cite{Kosowsky_Loeb96,Kahn10,Pogosian11}. 
However, the effects of a primordial magnetic field on the CMB 
are relatively more pronounced in its non-Gaussian correlations.
This arises due to the fact that magnetic fields induce non-Gaussian signals at lowest order 
as the magnetic energy density and stress are quadratic in the field. 
In contrast, the standard inflationary perturbations, 
dominated by their linear component, can source non-Gaussian correlations 
only with higher order perturbations and thus necessarily can only produce a small amplitude of 
CMB non-Gaussianity (cf. \cite{Verde00,Maldacena03,Bartolo04,Komatsu10,Crem04,Seery05,Lyth05,ByrnesSasakiWands06,komatsu_spergel01}).
Primordial magnetic fields have been shown capable of inducing appreciable CMB non-Gaussianity when considering the bispectrum 
\cite{SS09,Caprini09,Cai10,Shiraishi_vector,Shiraishi_tensor,Shiraishi_STautocross,Shiraishi_tensor_WMAP7,Shiraishi_pol,BC05,Brown11,TSS10}. 
Our earlier calculation of the magnetic CMB bispectrum sourced by 
scalar anisotropic stress led to a 
$\sim 2$ nG upper limit on the primordial magnetic field's amplitude on megaparsec scales \cite{TSS10}.
However, higher-order measures of non-Gaussianity like the trispectrum have been less investigated and
as we show here, are very useful to set further constraints on primordial magnetic fields.

In this article we present in detail the primordial magnetic field contribution to the CMB scalar mode trispectrum. 
The principal results were summarized in our earlier Letter \cite{TSS11_Trispec_Letter}, where WMAP5 and WMAP7 
constraints on non-Gaussianity were used to derive magnetic field constraints. Here we present the full trispectrum calculations
as well as an additional flat-sky calculation for the scalar anisotropic stress trispectrum. Furthermore, the new constraints 
on non-Gaussianity from the {\it Planck} mission 2013 data release \cite{Planck_NG} are utilized to obtain improved magnetic field constraints. 
We find that the trispectrum does better than the bispectrum at probing magnetic fields on large scales.We also show that even stronger constraints can be
imposed on magnetic fields by considering the recently discussed magnetic inflationary curvature mode \cite{bonvin13}.
  
In the next section we describe the properties of the stochastic primordial magnetic field assumed for our calculations. 
The Sachs-Wolfe effect sourced by the magnetic energy density of a stochastic primordial magnetic field is presented in Sec.~\ref{s:OmegaB}.  
The full mode-coupling calculations are then presented for the four-point correlation of 
magnetic energy density. In Sec.~\ref{s:PiB} we present the Sachs-Wolfe effect and four-point calculation for magnetic scalar anisotropic stress. 
The magnetic CMB trispectrum is then calculated for energy density and scalar anisotropic stress in Sec.~\ref{s:MagTrispec}.
Additionally, in Sec.~\ref{s:FlatSky}, the trispectrum sourced by magnetic scalar anisotropic stress is also calculated using the 
flat-sky approximation. 
Finally, in Sec.~\ref{s:BLimits}, 
the {\it Planck} 2013 data release constraints on CMB non-Gaussianity \cite{Planck_NG}
are used to place improved upper limits 
on the strength of primordial magnetic fields.

%+++++++++++++++++++++++++++++++++++++++++++++++++++++++++++++++++++++++

\section{Primordial Magnetic Field}
\label{s:PMF}

We consider a Gaussian random stochastic magnetic field ${\bf B}$ characterized and completely specified 
by its 
power spectrum $M(k)$. We further assume that the magnetic field is non-helical. On scales that are galactic 
and larger, any velocity induced by Lorentz forces is generally too small to 
appreciably distort the initial magnetic field \cite{Jedam98,SB98}.  Therefore, the magnetic field 
simply redshifts away as ${\bf B}({\bf x},t)={\bb}_{0}({\bf x})/a^{2}$, where, ${\bb}_{0}$ 
is the magnetic field at the present epoch (i.e. at $z=0$ or $a=1$).
We define ${\bb} ({\kk})$ as the Fourier transform of the magnetic field ${\bb}_0 ({\xx})$. 
The magnetic field power spectrum is defined as
\EQ
\langle b_{i}({\kk})b^*_{j}({\qq})\rangle=(2\pi)^3 \delta({{\kk}-{\qq}})P_{ij}({\kk})M(k)
\EN 
where $P_{ij}({\kk}) = (\delta_{ij} - k_ik_j/k^2)$ is the projection operator ensuring 
${\bf \nabla}\cdot{\bb_0} =0$. 
This gives  $\langle {\bb}_{0}^{2} \rangle=2\int (dk/k)\Delta _{b}^{2}(k)$, 
where $\Delta _{b}^{2}(k)=k^{3}M(k)/(2\pi ^{2})$ 
is the power per logarithmic interval in $k$-space present in the stochastic magnetic field.
We also assume a power-law magnetic power spectrum,
$M(k)=Ak^{n}$ that is cutoff at $k=k_{c}$,
where $k_{c}$ is the Alfv\'{e}n-wave damping length-scale \cite{Jedam98,SB98}. 
We then fix the normalization $A$ by setting the variance of the magnetic field to be $B_0$, 
smoothed using a sharp $k$-space filter, over a `galactic' scale $k_G=1h$ Mpc$^{-1}$. 
This gives, (for $n \gtrsim -3$ and for $k<k_c$)
\EQ
\Delta _{b}^{2}(k)= \frac{k^3M(k)}{2\pi^2}
=\frac{B_0^2}{2}(n+3)\left(\frac{k}{k_{G}}\right)^{3+n}.
\EN
We restrict the magnetic spectral index to values near and above -3, i.e an inflation-generated field,
as causal generation mechanisms necessarily produce much bluer magnetic power spectra \cite{durrer_caprini03}.
Furthermore, blue spectral indices, on large scales, are strongly disfavoured by many observational constraints 
on primordial magnetic fields like the CMB power spectra \cite{Planck_Parameters,Pao12,Pao10,Yam10,SL10}.

% 
%+++++++++++++++++++++++++++++++++++++++++++++++++++++++++++++++++++++++

\section{CMB Anisotropy from Magnetic Energy Density and Four-Point Correlation}
\label{s:OmegaB}

The Sachs-Wolfe type of contribution to the CMB temperature anisotropy sourced by 
the energy density of 
magnetic fields \cite{giovannini07_PMC,finelli08,bonvin10}, 
can be written as
\EQ
\frac{\Delta T}{T}(\nn) =  
{\cal R} ~ \Omega_B(\xx_0 -\nn D^*).
\EN
Here, $\Omega_B({\xx}) = {\bf B}^2({\xx},t)/(8\pi \rho_\gamma(t)) 
={\bb}_0^2({\xx})/(8\pi \rho_0)$, 
where $\rho_\gamma(t)$ and $\rho_0$ are the CMB energy densities at times $t$ and 
at the present epoch, respectively. Like the usual Sachs-Wolfe effect, the $\Delta T/T$ 
given above is for large-angular scales.  
For calculating numerical values we adopt the ${\cal R}$ value estimated by Bonvin and Caprini
(Eq. 6.12 of \cite{bonvin10}) which is expressed according to our definitions as 
${\cal R} = - R_\gamma/15 \sim -0.04$, where $R_\gamma \sim 0.6$ is the fractional 
contribution of radiation energy density towards the total energy density of the relativistic 
component. The unit vector ${\bf n}$ is defined along the direction of observation from the observer 
at position $\xx_0$ and $D^*$ is the 
(comoving angular diameter) distance to the surface of last scattering. We have assumed instantaneous 
recombination which is a good approximation for large angular scales.

The temperature fluctuations of the CMB can be expanded in terms of spherical harmonics to give 
$\Delta T(\nn)/T = \sum_{l m} a_{lm} Y_{lm}(\nn)$,
where
\EQ
a_{lm}= \frac{4 \pi}{i^l}\int \frac{d^3 k}{(2\pi)^3} ~
{\cal R} ~ \Omega_B(\kk) ~ j_l(kD^*) ~ Y^*_{lm}(\kkk). 
\label{alm}
\EN
Note that $\Omega_B(\kk)$ is the Fourier transform of $\Omega_B({\xx})$. 
As $\Omega_B({\xx})$ is quadratic in ${\bb}_0({\xx})$,  $\Omega_B(\kk)$ is given by the convolution integral
\EQ
\Omega_B(\kk) = \left({1}/{(2\pi)^{3}}\right)\int d^3s ~b_i(\kk+\ssvec)b^*_i(\ssvec)/(8\pi \rho_0).
\label{Omega_B_k_convolution}
\EN
The trispectrum $T^{m_{_1}\!m_{_2}\!m_{_3}\!m_{_4}}_{\,\,\,\,l_{_1}\,\;l_{_2}\,\;l_{_3}\,\;l_{_4}}$, 
or the four-point correlation function of the CMB temperature anisotropy in harmonic space, in terms of the ${a_{lm}}$'s is
\EQ
T^{m_{_1}\!m_{_2}\!m_{_3}\!m_{_4}}_{\,\,\,\,l_{_1}\,\;l_{_2}\,\;l_{_3}\,\;l_{_4}}
=\,\langle a_{{l_1}{m_1}}a_{{l_2}{m_2}}a_{{l_3}{m_3}}a_{{l_4}{m_4}}\rangle.
\EN
From Eq.(\ref{alm}) we can express $T^{m_{_1}\!m_{_2}\!m_{_3}\!m_{_4}}_{\,\,\,\,l_{_1}\,\;l_{_2}\,\;l_{_3}\,\;l_{_4}}$ as
\EQ
\!\!T^{m_{_1}\!m_{_2}\!m_{_3}\!m_{_4}}_{\,\,\,\,l_{_1}\,\;l_{_2}\,\;l_{_3}\,\;l_{_4}}
\!= \! \left(\!\frac{\cal R}{2\pi^2}\!\right)^{\!\!4}\!\!\int 
\!\left[\prod_{i=1}^4
\frac{d^3 k_i}{i^{l_i}} j_{_{l_i}}\!(k_{_i}D^*)Y^*_{l_im_i}\!(\hat{\kk}_{_i})\right]
\!\zeta_{_{1234}}
\label{trispec}
\EN
with 
\EQ
\zeta_{_{1234}}=\, \langle \Omega_B(\kk_{_1})\Omega_B(\kk_{_2})
\Omega_B(\kk_{_3})\Omega_B(\kk_{_4})\rangle.
\EN 
The four-point correlation function of $\Omega_B(\kk)$ involves an eight-point 
correlation function of the magnetic fields. Using Wick's Theorem, for Gaussian magnetic fields, 
we can express the magnetic eight-point correlation as a sum of 105 terms containing the magnetic 
two-point correlation. Neglecting 45 terms proportional to $\delta(\kk)$ that vanish and  12 terms proportional to $\delta(\kk_i + \kk_j)$
that are the unconnected part of the four-point correlation, 
48 terms remain. A long calculation using the relevant projection operators gives
$\zeta_{_{1234}} = 
\delta(\kk_1 + \kk_2 + \kk_3 + \kk_4) ~ \psi_{_{1234}}$, where $\psi_{_{1234}}$ is a mode-coupling
integral over a variable $\ssvec$ and also contains angular terms.

The full expression for $\psi_{_{1234}}$ involving angular terms in the 
mode-coupling integral is

\begin{widetext}
\EQA
\psi_{_{1234}} = \frac{8}{(8\pi\rho_0)^4}
\int d^3 s M(s) M(\vert \kk_1 + \ssvec \vert) \Big[ 
  &M&(\left| \kk_1 + \kk_3 +\ssvec \right|) \left( M(\vert \kk_2  -  \ssvec \vert) {\cal F}_{(1)} 
                                                + M(\vert \kk_4  -  \ssvec \vert) {\cal F}_{(2)} \right) \nonumber \\
+ &M&(\left| \kk_1 + \kk_2 +\ssvec \right|) \left( M(\vert \kk_3  -  \ssvec \vert) {\cal F}_{(3)} 
                                                + M(\vert \kk_4  -  \ssvec \vert) {\cal F}_{(4)} \right) \nonumber \\
+ &M&(\left| \kk_1 + \kk_4 +\ssvec \right|) \left( M(\vert \kk_2  -  \ssvec \vert) {\cal F}_{(5)} 
                                                + M(\vert \kk_3  -  \ssvec \vert) {\cal F}_{(6)} \right) \Big] 
\label{psi}
\ENA
with 
\EQA
 \mathcal{F}_{(1)}&=& -1 + \left( \alpha_1^2 + \alpha_2^2 +\alpha_6^2 + \beta_2^2 + \beta_6^2 + \gamma_6^2 \right)
     - \left( \alpha_1\alpha_2\beta_2 + \alpha_1\alpha_6\beta_6 + \alpha_2\alpha_6\gamma_6 + \beta_2\beta_6\gamma_6 \right)
     + \alpha_1\alpha_2\beta_6\gamma_6 \nonumber \\
 \mathcal{F}_{(2)}&=& -1 + \left( \alpha_1^2 + \alpha_4^2 +\alpha_6^2 + \beta_4^2 + \beta_6^2 + \epsilon_6^2 \right)
     - \left( \alpha_1\alpha_4\beta_4 + \alpha_1\alpha_6\beta_6 + \alpha_4\alpha_6\epsilon_6 + \beta_4\beta_6\epsilon_6 \right)
     + \alpha_1\alpha_4\beta_6\epsilon_6 \nonumber \\
 \mathcal{F}_{(3)}&=& -1 + \left( \alpha_1^2 + \alpha_3^2 +\alpha_5^2 + \beta_3^2 + \beta_5^2 + \delta_5^2 \right)
     - \left( \alpha_1\alpha_3\beta_3 + \alpha_1\alpha_5\beta_5 + \alpha_3\alpha_5\delta_5 + \beta_3\beta_5\delta_5 \right)
     + \alpha_1\alpha_3\beta_5\delta_5 \nonumber \\
 \mathcal{F}_{(4)}&=& -1 + \left( \alpha_1^2 + \alpha_4^2 +\alpha_5^2 + \beta_4^2 + \beta_5^2 + \epsilon_5^2 \right)
     - \left( \alpha_1\alpha_4\beta_4 + \alpha_1\alpha_5\beta_5 + \alpha_4\alpha_5\epsilon_5 + \beta_4\beta_5\epsilon_5 \right)
     + \alpha_1\alpha_4\beta_5\epsilon_5 \nonumber \\
 \mathcal{F}_{(5)}&=& -1 + \left( \alpha_1^2 + \alpha_2^2 +\alpha_7^2 + \beta_2^2 + \beta_7^2 + \gamma_7^2 \right)
     - \left( \alpha_1\alpha_2\beta_2 + \alpha_1\alpha_7\beta_7 + \alpha_2\alpha_7\gamma_7 + \beta_2\beta_7\gamma_7 \right)
     + \alpha_1\alpha_2\beta_7\gamma_7 \nonumber \\
 \mathcal{F}_{(6)}&=& -1 + \left( \alpha_1^2 + \alpha_3^2 +\alpha_7^2 + \beta_3^2 + \beta_7^2 + \delta_7^2 \right)
     - \left( \alpha_1\alpha_3\beta_3 + \alpha_1\alpha_7\beta_7 + \alpha_3\alpha_7\delta_7 + \beta_3\beta_7\delta_7 \right)
     + \alpha_1\alpha_3\beta_7\delta_7.
 \label{72terms}
\ENA

\begin{figure}[tbp]
\centering
\epsfig{file=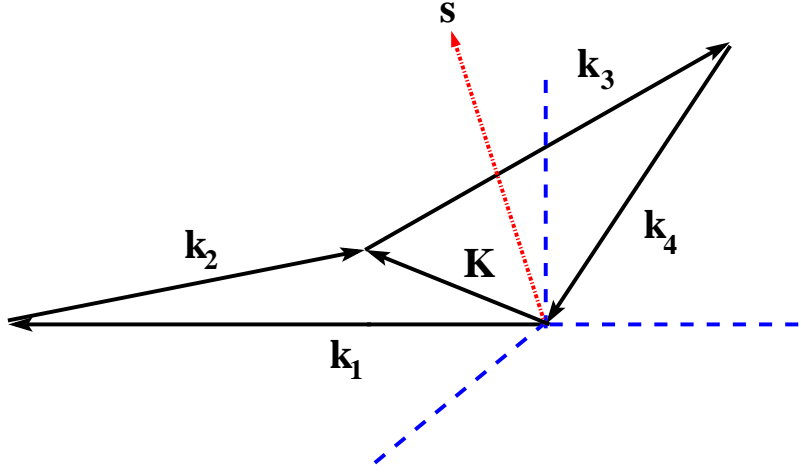,width=0.6\linewidth,clip=}
\caption{The general configuration of four wavevectors $\kk_1, \kk_2, \kk_3$ and $\kk_4$ for the trispectrum 
with the integration mode wavevector $\ssvec$ that appears in the mode-coupling integral.}
\label{fig_s_integration}
\end{figure}

\end{widetext}

The angular terms ${\mathcal F}$ contain angles defined according to 
\EQA
\phi_1 = &\wwhat& \cdot \widehat{\kk_1 + \ssvec}, \, \, 
\phi_2 = \wwhat \cdot \widehat{\kk_2 - \ssvec}, \, \,
\phi_3 = \wwhat \cdot \widehat{\kk_3 - \ssvec}, \nonumber \\
\phi_4 = &\wwhat& \cdot \widehat{\kk_4 - \ssvec}, \, \,
\phi_5 = \wwhat \cdot \widehat{\overline{\kk_1 + \kk_2 + \ssvec}}, \nonumber \\
\phi_6 = &\wwhat& \cdot \widehat{\overline{\kk_1 + \kk_3 + \ssvec}}, \, \,
\phi_7 = \wwhat \cdot \widehat{\overline{\kk_1 + \kk_4 + \ssvec}},
\label{phi_subscript_def}
\ENA
where $\widehat{\kk_1 + \ssvec}$ is a unit vector in the direction of $(\kk_1 + \ssvec)$ and 
the angle $\phi$ denotes different angles for different values of the unit vector $\wwhat$  
\EQA
\phi &=& \alpha \: \text{ for } \: \wwhat = \sss, \; \; \;
\phi = \beta \: \text{ for } \: \wwhat = \widehat{\kk_1 + \ssvec}, \nonumber \\
\phi &=& \gamma \: \text{ for } \: \wwhat = \widehat{\kk_2 - \ssvec}, \; \; \;
\phi = \delta \: \text{ for } \: \wwhat = \widehat{\kk_3 - \ssvec}, \nonumber \\
\phi &=& \epsilon \: \text{ for } \: \wwhat = \widehat{\kk_4 - \ssvec}, \; \; \;
\phi = \kappa \: \text{ for } \: \wwhat = \widehat{\overline{\kk_1 + \kk_2 + \ssvec}}, \nonumber \\
\phi &=& \lambda \: \text{ for } \: \wwhat = \widehat{\overline{\kk_1 + \kk_3 + \ssvec}}.
\label{phi_def}
\ENA

For simplicity of calculation we evaluate the mode-coupling integral $\psi_{_{1234}}$ in two cases: 
(I) considering only $\ssvec$-independent angular terms for all equal-sided configurations and 
(II) taking all angular terms for the collinear configuration. 

\subsection{Case I - $\ssvec$-independent terms for equal-sided configurations}

Considering only $\ssvec$-independent angular terms, for a general configuration, we find
$\psi_{_{1234}} = {-8}/{(8\pi\rho_0)^4} ~ {\cal I}$
where
\EQA
{\cal I} &=& \int d^3 s \; M(s) \; M(\vert \kk_1 + \ssvec \vert) \times \nonumber \\ 
  \Big[ &M& (\left| \kk_1 + \kk_3 +\ssvec \right|) \Big( M(\vert \kk_2  -  \ssvec \vert) 
                                                + M(\vert \kk_4  -  \ssvec \vert) \Big) \nonumber \\
+ &M&(\left| \kk_1 + \kk_2 +\ssvec \right|) \Big( M(\vert \kk_3  -  \ssvec \vert)  
                                                + M(\vert \kk_4  -  \ssvec \vert) \Big) \nonumber \\
+ &M&(\left| \kk_1 + \kk_4 +\ssvec \right|) \Big( M(\vert \kk_2  -  \ssvec \vert) 
                                                + M(\vert \kk_3  -  \ssvec \vert) \Big) \Big] \nonumber \\
&=& \; \; {\cal I}_{\rm (1)} +  {\cal I}_{\rm (2)} + {\cal I}_{\rm (3)} + {\cal I}_{\rm (4)} 
+ {\cal I}_{\rm (5)} + {\cal I}_{\rm (6)}.
\label{calI_I}
\ENA
We perform the mode-coupling integral employing the technique and approximations 
discussed in \cite{TSS10,trsks01,mack02,SSB03}, while adopting the mean (zero) value of $\kkk_1 \cdot \kkk_3$, to find, 
for the first term,
\EQ
{\cal I}_{\rm (1)} \simeq 4 \pi A^4 ~ k_1^{2n+3} ~ k_2^n ~ k_3^n \left[ \frac{2^{n/2}}{n+3} - \frac{1}{4n+3} \right].
\label{Ii}
\EN
The value of each of the ${\mathcal I}_{(j)}$ integrals for $j={\rm 1}$ to ${\rm 6}$ is the same 
when all the $\kk_i$ wavevectors are of equal magnitude $\vert \kk_i \vert = k$. 
We perform the $s$-independent (case I) trispectrum evaluation for such equal-sided quadrilateral configurations. 
Hence, ${\mathcal I}= \sum_{j=(1)}^{(6)} 
{\mathcal I}_j 
= 6 \,\, {\mathcal I}_{\rm (1)}$, and 
we obtain
\EQA
&\!&\zeta_{_{1234}} \; = \;\;\delta(\kk_1 + \kk_2 + \kk_3 + \kk_4) \times \nonumber \\
&&\frac{-8 \, (24\pi)\, A^4 \,k_1^{2n+3} \,k_2^n \,k_3^n }{(8\pi\rho_0)^4}\left[ \frac{(2^{n/2})(4n+3)-(n+3)}{(4n+3)(n+3)} \right]\!.
\label{zeta_Omega}
\ENA

\subsection{Case II - Equal-Sided Collinear Configuration}

We calculate the full mode-coupling integral $\psi_{_{1234}}$ (Eq. \ref{psi},\ref{72terms}) (over all angular terms 
for each ${\mathcal F}$ expression) for the case of the equal-sided collinear configuration. All the 
four wavevectors are of equal magnitude with configuration $\kk_1 = \kk_2 = -\kk_3 = -\kk_4$. We find that 
the 28 independent angles defined by Equations (\ref{phi_subscript_def},\ref{phi_def}) reduce to just 6 independent
angles $\alpha_1,\alpha_2,\alpha_5,\beta_2,\beta_5 \text{ and } \gamma_5$. The angular expressions 
${\mathcal F}$ also reduce in size from a total of 72 to 19 angular terms:
\EQ
\psi^{\text{coll}}_{_{1234}} = \frac{8}{(8\pi\rho_0)^4} ~ {\cal I}^{\text{coll}}
\label{psi_II}
\EN
where
\EQA
{\cal I}^{\text{coll}} &=& 2 \, \int d^3 s \; M(s) \; M(\vert \kk + \ssvec \vert) \times \nonumber \\ 
  \Big[ &M&(s) M(\left| \kk - \ssvec \right|) 
       \Big( \alpha_1^2 + \alpha_2^2 + \beta_2^2 - 2 \alpha_1\alpha_2\beta_2 + \alpha_1^2 \alpha_2^2 \Big) \nonumber \\
+ &M&(s) M(\left| \kk + \ssvec \right|) 
       \Big( 1 + \alpha_4^2 \Big) \nonumber \\
+ &M&(\left| 2\kk + \ssvec \right|) M(\left| \kk + \ssvec \right|) 
       \Big( \alpha_1^2 + \alpha_5^2 + \beta_5^2 - \alpha_1\alpha_5\beta_5 \nonumber \\
             && + \frac{1}{2} \left \{ \delta_5^2 + \epsilon_5^2 + \left( \beta_5 + \alpha_1\alpha_5 - \alpha_1^2 \beta_5 \right) 
             \left( \delta_5 + \epsilon_5 \right) \right \} \Big) \Big]. \nonumber \\
\label{calI_II}
\ENA
Using the same technique of evaluating the mode-coupling integrals as used earlier in Case I, we calculate the integrals 
for each of the 19 angular terms that sum together to give
\EQ
{\cal I}^{\text{coll}} \simeq4 \pi A^4 k_1^{2n+3} k_2^n k_3^n \left[ \frac{8}{3}\frac{2^{n/2}}{n+3} - \frac{12}{4n+3} \right].
\label{Iii}
\EN
The four-point correlation of magnetic energy density for the collinear configuration is 
\EQA
&\zeta_{_{1234}}& \; = \;\;\delta(\kk_1 + \kk_2 + \kk_3 + \kk_4) \times \nonumber \\
&\!&\!\!\!\!\!\!\!\!\!\!\!\!\!\!\!\!\!\!\frac{8 \, (4\pi)\, A^4 \,k_1^{2n+3} k_2^n k_3^n }{(8\pi\rho_0)^4}
\!\!\left[ \frac{\frac{8}{3}(2^{n/2})(4n+3)-(12)(n+3)}{(4n+3)(n+3)} \right]\!\!.
\label{zeta_coll}
\ENA

%+++++++++++++++++++++++++++++++++++++++++++++++++++++++++++++++++++++++

\section{CMB Anisotropy from Magnetic Scalar Anisotropic Stress and Four-Point Correlation}
\label{s:PiB}

The scalar anisotropic stress that is associated with a primordial magnetic field, in addition to its energy density, 
will also act as a separate source for CMB fluctuations - the passive mode \cite{SL09,bonvin10}. 
As we saw in our previous work \cite{TSS10}, the magnetic scalar anisotropic stress generates $\sim 10^6$ times 
larger contribution to the CMB bispectrum compared to magnetic energy density. With this motivation in mind and
employing the magnetic CMB trispectrum technique developed above, we carry out a longer calculation 
for the scalar anisotropic stress trispectrum. 

On large angular scales, the magnetic contribution to the temperature anisotropy is again via the magnetic Sachs-Wolfe effect 
\EQ
\frac{\Delta T}{T}(\nn) = \frac{1}{3} \, \Phi(\xx_0 -\nn D^*) = \frac{1}{5} \, \zeta(\xx_0 -\nn D^*)
\label{SW_Phi_Zeta}
\EN
in the matter dominated era. We use the expression for the curvature perturbation 
due to the passive mode scalar anisotropic stress \cite{SL09} 
\EQ
\zeta \simeq  - \frac{1}{3} R_{\gamma} \Pi_B \ln \left( \frac{\tau_{\nu}}{\tau_B}\right).
\label{zeta}
\EN
to obtain temperature anisotropy, sourced by magnetic scalar anisotropic stress $\Pi_B$
\EQ
\frac{\Delta T}{T}(\nn) =  
\mathcal{R}_p ~ \Pi_B(\xx_0 -\nn D^*),
\label{Delta_T_stress}
\EN
where $\mathcal{R}_p 
= \mathcal{R} ~\ln \left( {\tau_{\nu}}/{\tau_B} \right) 
= \left[ -R_{\gamma}/15 \right] ~\ln \left( {T_{B}}/{T_{\nu}} \right)$ and 
 $\tau_B$ as well as $\tau_{\nu}$ and 
$T_B$ as well as $T_{\nu}$ are the 
conformal time and temperatures at the epochs of 
magnetic field generation and neutrino decoupling, respectively.
None of the details of the magnetic scalar anisotropic stress calculation were included in our letter \cite{TSS11_Trispec_Letter} 
and they are presented below.

The CMB temperature fluctuations can be expanded in terms of spherical harmonics to give 
$\Delta T(\nn)/T = \sum_{l m} a_{lm} Y_{lm}(\nn)$,
where
\EQ
a_{lm}= \frac{4 \pi}{i^l}\int \frac{d^3 k}{(2\pi)^3} ~
{\cal R}_p ~ \Pi_B(\kk) ~ j_l(kD^*) ~ Y^*_{lm}(\kkk). 
\label{alm_Pi}
\EN
Here, $\Pi_B(\kk)$ is the Fourier transform of $\Pi_B({\xx})$ and we recall the operator that 
projects out the scalar anisotropic stress from the full magnetic stress $\Pi_B^{ij}(\kk)$
\EQ
\Pi_B(\kk) = \frac{1}{2} \left( \delta_{ij} - 3 \kkk _i \kkk_j \right) \Pi_B^{ij}(\kk)
\label{Pi_k}
\EN
Since $\Pi_B({\xx})$ is quadratic in ${\bb}_0({\xx})$, we have a convolution of magnetic fields
% (*** Ummm ***)\\
\EQ 
\Pi_B(\kk) = \frac{1}{2} \left( \delta_{ij} - 3 \kkk _i \kkk_j \right) \frac{1}{4 \pi p_{\gamma}}
\int \frac{d^3 s}{(2\pi)^{3}} ~\bb^*_i(\ssvec) \bb_j(\kk+\ssvec)
\label{Pi_k_s_integral}
\EN
The trispectrum 
is
$T^{m_{_1}\!m_{_2}\!m_{_3}\!m_{_4}}_{\,\,\,\,l_{_1}\,\;l_{_2}\,\;l_{_3}\,\;l_{_4}}
=\,\langle a_{{l_1}{m_1}}a_{{l_2}{m_2}}a_{{l_3}{m_3}}a_{{l_4}{m_4}}\rangle$, is then given by 
\EQ
\!\!T^{m_{_1}\!m_{_2}\!m_{_3}\!m_{_4}}_{\,\,\,\,l_{_1}\,\;l_{_2}\,\;l_{_3}\,\;l_{_4}}
\!= \! \left(\!\frac{{\cal R}_p}{2\pi^2}\!\right)^{\!\!4}\!\!\int 
\!\left[\prod_{i=1}^4
\frac{d^3 k_i}{i^{l_i}} j_{_{l_i}}\!(k_{_i}D^*)Y^*_{l_im_i}\!(\hat{\kk}_{_i})\right]
\!\left[ \zeta_{_{1234}} \right]_{\Pi}
\label{trispec_Pi}
\EN
with 
$\left[ \zeta_{_{1234}}\right]_{\Pi}$ defined as
\EQ
\left[ \zeta_{_{1234}}\right]_{\Pi}=\, \langle \Pi_B(\kk_{_1})\Pi_B(\kk_{_2})
\Pi_B(\kk_{_3})\Pi_B(\kk_{_4})\rangle.
\label{zeta_Pi_definition}
\EN 
The four-point correlation function of $\Pi_B(\kk)$, like that of $\Omega_B(\kk)$, also involves an eight-point 
correlation function of the fields. In similar fashion, using Wick's Theorem, for Gaussian magnetic fields, 
we express the magnetic eight-point correlation as a sum of 105 terms involving the magnetic 
two-point correlation function. Then 45 terms proportional to $\delta(\kk)$ vanish and we neglect 
the 12 terms proportional to $\delta(\kk_i + \kk_j)$ that represent the unconnected part of the four-point correlation, 
to leave 48 terms. A long calculation involving the relevant projection operators in these terms gives
$\left[ \zeta_{_{1234}}\right]_{\Pi} = \delta(\kk_1 + \kk_2 + \kk_3 + \kk_4) ~ \left[ \psi_{_{1234}}\right]_{\Pi}$, 
where $\left[ \psi_{_{1234}} \right]_{\Pi}$ is a mode-coupling
integral over a variable $\ssvec$ and also involves angular terms. The key difference between the $\Omega_B$ and the $\Pi_B$ 
four-point correlations is the number and type of operators acting on the magnetic field eight-point correlation. 
In the case of energy density $\Omega_B$, the operator $\delta_{ab}\delta_{cd}\delta_{ef}\delta_{gh}$ acted on
\begin{widetext} 
\EQ
\langle \bb_a(-\ssvec) \bb_b(\kk_1 + \ssvec) \bb_c(-\rr) \bb_d(\kk_2 + \rr) \bb_e(-\ttvec) \bb_f(\kk_3 + \ttvec) 
\bb_g(-\wwvec) \bb_h(\kk_4 + \wwvec) \rangle.
\EN
However, in the case of scalar anisotropic stress $\Pi_B$, there are 16 operator terms
\EQA
\left( \delta_{_{ab}}\!\! \right. && \left. - 3 \kkk_{1_{a}} \kkk_{1_{b}} \!\right)
\left( \delta_{_{cd}}\!\! - 3 \kkk_{2_{c}} \kkk_{2_{d}} \!\right)
\left( \delta_{_{ef}}\!\! - 3 \kkk_{3_{e}} \kkk_{3_{f}} \!\right)
\left( \delta_{_{gh}}\!\! - 3 \kkk_{4_{g}} \kkk_{4_{h}} \!\right) \nonumber \\
= && \; \; \delta_{_{ab}} \delta_{_{cd}}\delta_{_{ef}}\delta_{_{gh}}
     - \; 3 \left[    \delta_{_{ab}}\delta_{_{cd}}\delta_{_{ef}}\kkk_{4_{g}}\kkk_{4_{h}}
                     +\delta_{_{ab}}\delta_{_{cd}}\kkk_{3_{e}}\kkk_{3_{f}}\delta_{_{gh}}
                     +\delta_{_{ab}}\kkk_{2_{c}}\kkk_{2_{d}}\delta_{_{ef}}\delta_{_{gh}}
                     +\kkk_{1_{a}}\kkk_{1_{b}}\delta_{_{cd}}\delta_{_{ef}}\delta_{_{gh}} \right] \nonumber \\
  && + \; 9 \left[    \delta_{_{ab}}\delta_{_{cd}}\kkk_{3_{e}}\kkk_{3_{f}}\kkk_{4_{g}}\kkk_{4_{h}}
                     +\delta_{_{ab}}\kkk_{2_{c}}\kkk_{2_{d}}\delta_{_{ef}}\kkk_{4_{g}}\kkk_{4_{h}}
                     +\delta_{_{ab}}\kkk_{2_{c}}\kkk_{2_{d}}\kkk_{3_{e}}\kkk_{3_{f}}\delta_{_{gh}}
                     +\kkk_{1_{a}}\kkk_{1_{b}}\delta_{_{cd}}\delta_{_{ef}}\kkk_{4_{g}}\kkk_{4_{h}} \right. \nonumber \\
  && \left. \;\;\;\;\;
                     +\kkk_{1_{a}}\kkk_{1_{b}}\delta_{_{cd}}\kkk_{3_{e}}\kkk_{3_{f}}\delta_{_{gh}}
                     +\kkk_{1_{a}}\kkk_{1_{b}}\kkk_{2_{c}}\kkk_{2_{d}}\delta_{_{ef}}\delta_{_{gh}}
                 \right] \nonumber \\
 && - \; 27 \left[    \delta_{_{ab}}\kkk_{2_{c}}\kkk_{2_{d}}\kkk_{3_{e}}\kkk_{3_{f}}\kkk_{4_{g}}\kkk_{4_{h}}
                     +\kkk_{1_{a}}\kkk_{1_{b}}\delta_{_{cd}}\kkk_{3_{e}}\kkk_{3_{f}}\kkk_{4_{g}}\kkk_{4_{h}}
                     +\kkk_{1_{a}}\kkk_{1_{b}}\kkk_{2_{c}}\kkk_{2_{d}}\delta_{_{ef}}\kkk_{4_{g}}\kkk_{4_{h}}
                     +\kkk_{1_{a}}\kkk_{1_{b}}\kkk_{2_{c}}\kkk_{2_{d}}\kkk_{3_{e}}\kkk_{3_{f}}\delta_{_{gh}} \right] \nonumber \\
 && + \; 81 \;        \kkk_{1_{a}}\kkk_{1_{b}}\kkk_{2_{c}}\kkk_{2_{d}}\kkk_{3_{e}}\kkk_{3_{f}}\kkk_{4_{g}}\kkk_{4_{h}} \nonumber \\
= && \; \fbox{1} + \fbox{2} + ..... + \fbox{16}
\label{16operators_Pi}
\ENA
Each operator term $X$ from 1 to 16 generates its own separate angular term expression ${\cal F}_{(I)}^{\fbox{\tiny{X}}}$. When summed over all $X$ this yields
the angular term expression ${\cal F}_{(I)}$, where $I$ takes values 1 to 6 in the six term mode-coupling integral $\left[ \psi_{_{1234}}\right]_{\Pi}$. 
As operator $\fbox{1}$ is identical to the operator for the $\Omega_B$ four-point correlation, 
the angular terms ${\cal F}$ for it are
just given by 
Equation (\ref{72terms}). 
We give below the expressions for
$\left[ \psi_{_{1234}}\right]_{\Pi}$ and the angular terms ${\cal F}$ generated by operators $\fbox{2}$ and $\fbox{16}$, 
suppressing the $\Pi$ subscript. The complete expression for the full set of over 1500 angular terms generated by 
all sixteen operators $\fbox{1}$ through $\fbox{16}$ is placed in Appendix A. The mode-coupling integral for scalar anisotropic stress is,
\EQA
\left[ \psi_{_{1234}}\right]_{\Pi} = \frac{8}{(8\pi p_0)^4}
\int d^3 s M(s) M(\vert \kk_1 + \ssvec \vert) \Big[ 
  &M&(\left| \kk_1 + \kk_3 +\ssvec \right|) \left( M(\vert \kk_2  -  \ssvec \vert) {\cal F}_{(1)} 
                                                + M(\vert \kk_4  -  \ssvec \vert) {\cal F}_{(2)} \right) \nonumber \\
+ &M&(\left| \kk_1 + \kk_2 +\ssvec \right|) \left( M(\vert \kk_3  -  \ssvec \vert) {\cal F}_{(3)} 
                                                + M(\vert \kk_4  -  \ssvec \vert) {\cal F}_{(4)} \right) \nonumber \\
+ &M&(\left| \kk_1 + \kk_4 +\ssvec \right|) \left( M(\vert \kk_2  -  \ssvec \vert) {\cal F}_{(5)} 
                                                + M(\vert \kk_3  -  \ssvec \vert) {\cal F}_{(6)} \right) \Big] 
\label{psi_Pi_2}
\ENA
where $p_0 = \rho_0/3$ and $\rho_0$ is the present-day energy density in radiation. The angular expressions ${\cal F}$ now involve 
32 new angles (with overbars) defined below, in addition to the 28 previously defined angles (without overbars) 
that appear in the $\Omega_B$ expression - Equations (\ref{phi_subscript_def},\ref{phi_def}). 
The new angles defined in Table \ref{Pi_angles_overbars} arise from dot products 
of the four $\kkk$ wavevectors with the vector $\sss$ or with those combinations of $\ssvec$ and the four $\kkk$ wavevectors 
that appear in the equation for $\psi_{_{1234}}$.

%=========================TABLE===========================
\begin{table}
\caption{Angle definitions for scalar anisotropic stress $\Pi_B$ angular terms, with $i$= 1 to 4. }
\begin{tabular}{cccccccc} 
\hline \hline \\
% \multicolumn{1}{c}{Configuration} &
% % \multicolumn{1}{c}{$(\kk_1,\kk_2,\kk_3,\kk_4)$} &
% \multicolumn{1}{c}{$\quad (\theta_{12},\theta_{13},\theta_{14},\theta_{23},\theta_{24},\theta_{34})\quad$} &
% \multicolumn{1}{c}{$m$}
\multicolumn{1}{c}{$\overline{\alpha}_i$} &
\multicolumn{1}{c}{$\overline{\beta}_i$} &
\multicolumn{1}{c}{$\overline{\gamma}_i$} &
\multicolumn{1}{c}{$\overline{\delta}_i$} &
\multicolumn{1}{c}{$\overline{\epsilon}_i$} &
\multicolumn{1}{c}{$\overline{\kappa}_i$} &
\multicolumn{1}{c}{$\overline{\lambda}_i$} &
\multicolumn{1}{c}{$\overline{\chi}_i$}
 \\ [3 pt]
\hline
%_/_/_/_/_/_/_/_/_/_/_/_/_/_/_/_/_/_/_/_/_/_/_/_/_/_/_/_/_/_/_/_/
\\
$\kkk_i \cdot \sss$ \; \; \;&
$\kkk_i \cdot \widehat{\kk_1 + \ssvec}$ \; \; \; &
$\kkk_i \cdot \widehat{\kk_2 - \ssvec}$ \; \; \; &
$\kkk_i \cdot \widehat{\kk_3 - \ssvec}$ \; \; \; &
$\kkk_i \cdot \widehat{\kk_4 - \ssvec}$ \; \; \; &
$\kkk_i \cdot \widehat{\overline{\kk_1 + \kk_2 + \ssvec}}$ \; \; \; &
$\kkk_i \cdot \widehat{\overline{\kk_1 + \kk_3 + \ssvec}}$ \; \; \; &
$\kkk_i \cdot \widehat{\overline{\kk_1 + \kk_4 + \ssvec}}$ \\ [5 pt] 
%_/_/_/_/_/_/_/_/_/_/_/_/_/_/_/_/_/_/_/_/_/_/_/_/_/_/_/_/_/_/_/_/
\hline \hline
\end{tabular}
% \footnotetext[1]{For the squeezed collinear configuration case, $\mathcal
% F_{\Pi_B \Pi_B \Pi_B }^{\text{I}}$ picks up another term $\bar{\mu}^2 \sim 1$}.
\label{Pi_angles_overbars}
\end{table}

% ==============================================================
The angular terms for operator $\fbox{2}$ are
\EQA
 \mathcal{F}_{(1)}^{\fbox{\tiny{2}}}&=& 1 - \overline{\beta}_4^2 -\overline{\alpha}_4^2 + \alpha_1\overline{\alpha}_4\overline{\beta}_4
 - \overline{\lambda}_4\left[ \overline{\lambda}_4 - \overline{\beta}_4\beta_6 - \overline{\alpha}_4\alpha_6 + \alpha_1\overline{\alpha}_4\beta_6 \right]
 - \overline{\gamma}_4\left[ \overline{\gamma}_4 - \overline{\beta}_4\beta_2 - \overline{\alpha}_4\alpha_2 + \alpha_1\overline{\beta}_4\alpha_2 \right] \nonumber \\
&& + \; \overline{\gamma}_4\overline{\lambda}_4\left[ \gamma_6 - \beta_2\beta_6 - \alpha_2\alpha_6 + \alpha_1\alpha_2\beta_6 \right] \nonumber \\
%--------------------
 \mathcal{F}_{(2)}^{\fbox{\tiny{2}}}&=&  1 - \overline{\beta}_4^2 -\overline{\lambda}_4^2 + \beta_6\overline{\lambda}_4\overline{\beta}_4
 - \overline{\epsilon}_4\left[ \overline{\epsilon}_4 - \overline{\beta}_4\beta_6 - \overline{\lambda}_4\epsilon_6 + \beta_6\overline{\beta}_4\epsilon_6 \right]
 - \overline{\alpha}_4\left[ \overline{\alpha}_4 - \overline{\beta}_4\alpha_1 - \overline{\lambda}_4\alpha_6 + \beta_6\overline{\lambda}_4\alpha_1 \right] \nonumber \\
&& + \; \overline{\alpha}_4\overline{\epsilon}_4\left[ \alpha_4 - \alpha_1\beta_4 - \alpha_6\epsilon_6 + \beta_6\alpha_1\epsilon_6 \right] \nonumber \\
%--------------------
 \mathcal{F}_{(3)}^{\fbox{\tiny{2}}}&=& 1 - \overline{\beta}_4^2 -\overline{\alpha}_4^2 + \alpha_1\overline{\alpha}_4\overline{\beta}_4
 - \overline{\kappa}_4\left[ \overline{\kappa}_4 - \overline{\beta}_4\beta_5 - \overline{\alpha}_4\alpha_5 + \alpha_1\overline{\alpha}_4\beta_5 \right]
 - \overline{\delta}_4\left[ \overline{\delta}_4 - \overline{\beta}_4\beta_3 - \overline{\alpha}_4\alpha_3 + \alpha_1\overline{\beta}_4\alpha_3 \right] \nonumber \\
&& + \; \overline{\delta}_4\overline{\kappa}_4\left[ \delta_5 - \beta_3\beta_5 - \alpha_3\alpha_5 + \alpha_1\alpha_3\beta_5 \right] \nonumber \\
%--------------------
 \mathcal{F}_{(4)}^{\fbox{\tiny{2}}}&=& 1 - \overline{\beta}_4^2 -\overline{\kappa}_4^2 + \beta_5\overline{\kappa}_4\overline{\beta}_4
 - \overline{\epsilon}_4\left[ \overline{\epsilon}_4 - \overline{\kappa}_4\epsilon_5 - \overline{\beta_4}\beta_4 + \beta_5\overline{\beta}_4\epsilon_5 \right]
 - \overline{\alpha}_4\left[ \overline{\alpha}_4 - \overline{\kappa}_4\alpha_5 - \overline{\beta}_4\alpha_1 + \beta_5\overline{\kappa}_4\alpha_1 \right] \nonumber \\
&& + \; \overline{\alpha}_4\overline{\epsilon}_4\left[ \alpha_4 - \alpha_5\epsilon_5 - \alpha_1\beta_4 + \beta_5\alpha_1\epsilon_5 \right] \nonumber \\
%--------------------
 \mathcal{F}_{(5)}^{\fbox{\tiny{2}}}&=& 1 - \overline{\gamma}_4^2 -\overline{\alpha}_4^2 + \alpha_2\overline{\alpha}_4\overline{\gamma}_4
 - \overline{\chi}_4\left[ \overline{\chi}_4 - \overline{\gamma}_4\gamma_7 - \overline{\alpha}_4\alpha_7 + \alpha_2\overline{\alpha}_4\gamma_7 \right]
 - \overline{\beta}_4\left[ \overline{\beta}_4 - \overline{\gamma}_4\beta_2 - \overline{\alpha}_4\alpha_1 + \alpha_2\overline{\gamma}_4\alpha_1 \right] \nonumber \\
&& + \; \overline{\beta}_4\overline{\chi}_4\left[ \beta_7 - \beta_2\gamma_7 - \alpha_1\alpha_7 + \alpha_2\gamma_7\alpha_1 \right] \nonumber \\
%--------------------
 \mathcal{F}_{(6)}^{\fbox{\tiny{2}}}&=& 1 - \overline{\alpha}_4^2 -\overline{\delta}_4^2 + \alpha_3\overline{\alpha}_4\overline{\delta}_4
 - \overline{\chi}_4\left[ \overline{\chi}_4 - \overline{\alpha}_4\alpha_7 - \overline{\delta}_4\delta_7 + \alpha_3\overline{\alpha}_4\delta_7 \right]
 - \overline{\beta}_4\left[ \overline{\beta}_4 - \overline{\alpha}_4\alpha_1 - \overline{\delta}_4\beta_3 + \alpha_3\overline{\delta}_4\alpha_1 \right] \nonumber \\
&& + \; \overline{\beta}_4\overline{\chi}_4\left[ \beta_7 - \alpha_1\alpha_7 - \beta_3\delta_7 + \alpha_3\alpha_1\delta_7 \right]
\label{ang_terms_Pi_2}
\ENA
and angular terms for operator $\fbox{16}$ are
\EQA
 \mathcal{F}_{(1)}^{\fbox{\tiny{16}}}&=&  \left( \theta_{12} -\overline{\alpha}_1 \overline{\alpha}_2 \right) 
                                            \left( \theta_{13} -\overline{\beta}_1 \overline{\beta}_3 \right) 
                                             \left( \theta_{24} -\overline{\gamma}_2 \overline{\gamma}_4 \right) 
                                              \left( \theta_{34} -\overline{\lambda}_3 \overline{\lambda}_4 \right) \nonumber \\
 \mathcal{F}_{(2)}^{\fbox{\tiny{16}}}&=&  \left( \theta_{14} -\overline{\alpha}_1 \overline{\alpha}_4 \right)
                                            \left( \theta_{13} -\overline{\beta}_1 \overline{\beta}_3 \right)
                                             \left( \theta_{24} -\overline{\epsilon}_2 \overline{\epsilon}_4 \right) 
                                              \left( \theta_{23} -\overline{\lambda}_2 \overline{\lambda}_3 \right)  \nonumber \\
 \mathcal{F}_{(3)}^{\fbox{\tiny{16}}}&=&  \left( \theta_{13} -\overline{\alpha}_1 \overline{\alpha}_3 \right)
                                            \left( \theta_{12} -\overline{\beta}_1 \overline{\beta}_2 \right)
                                             \left( \theta_{34} -\overline{\delta}_3 \overline{\delta}_4 \right) 
                                              \left( \theta_{24} -\overline{\kappa}_2 \overline{\kappa}_4 \right)  \nonumber \\
 \mathcal{F}_{(4)}^{\fbox{\tiny{16}}}&=&  \left( \theta_{14} -\overline{\alpha}_1 \overline{\alpha}_4 \right)
                                            \left( \theta_{12} -\overline{\beta}_1 \overline{\beta}_2 \right)
                                             \left( \theta_{34} -\overline{\epsilon}_3 \overline{\epsilon}_4 \right) 
                                              \left( \theta_{23} -\overline{\kappa}_2 \overline{\kappa}_3 \right)  \nonumber \\
 \mathcal{F}_{(5)}^{\fbox{\tiny{16}}}&=&  \left( \theta_{12} -\overline{\alpha}_1 \overline{\alpha}_2 \right)
                                            \left( \theta_{14} -\overline{\beta}_1 \overline{\beta}_4 \right)
                                             \left( \theta_{23} -\overline{\gamma}_2 \overline{\gamma}_3 \right) 
                                              \left( \theta_{34} -\overline{\chi}_3 \overline{\chi}_4 \right)  \nonumber \\
 \mathcal{F}_{(6)}^{\fbox{\tiny{16}}}&=&  \left( \theta_{13} -\overline{\alpha}_1 \overline{\alpha}_3 \right)
                                            \left( \theta_{14} -\overline{\beta}_1 \overline{\beta}_4 \right)
                                             \left( \theta_{23} -\overline{\delta}_2 \overline{\delta}_3 \right) 
                                              \left( \theta_{24} -\overline{\chi}_2 \overline{\chi}_4 \right).   
\label{ang_terms_Pi_16}
\ENA

\end{widetext}

\begin{figure}[tbp]
\centering
\epsfig{file=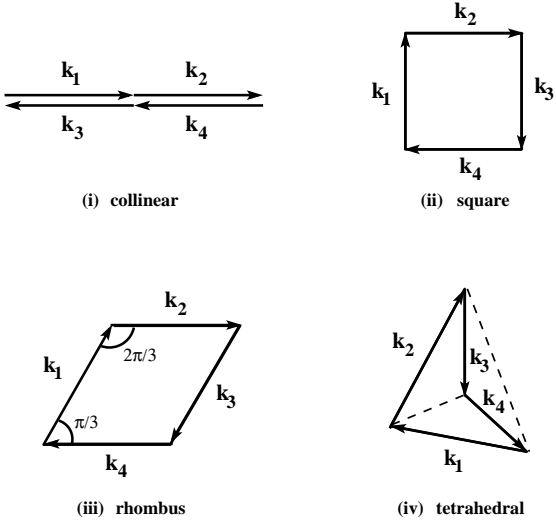,width=0.85\linewidth,clip=}
\caption{The four specific configurations (i) collinear, (ii) square, (iii) rhombus and (iv) tetrahedral, 
with each wavevector of equal magnitude $k$, used to evaluate the magnetic scalar anisotropic trispectrum.}
\label{fig_4_configs}
\end{figure}

In addition to the angles defined in Table \ref{Pi_angles_overbars}, angular terms like $\theta_{ab} = \kkk_a \cdot \kkk_b$ 
that are constant for a given $(\kk_1,\kk_2,\kk_3,\kk_4)$ configuration also appear. In total, 
as pointed out above, over 1500 angular terms 
are present in all the $\cal F$ expressions for $\left[ \psi_{_{1234}}\right]_{\Pi}$, many more than the 72 terms for
$\left[ \psi_{_{1234}}\right]_{\Omega}$. To arrive at a representative estimate for $\left[ \psi_{_{1234}}\right]_{\Pi}$, 
we consider only the $s$-independent angular terms and restrict ourselves to equal-sided trispectrum configurations i.e. all 
$\vert \kk_i \vert \simeq k$. The $s$-independent terms are
\EQA
&&\mathcal{F}^{\text{s-indep}}_{\Pi} = 6~\left[ -13 
+ 9 \left( \theta_{12}^2 + \theta_{13}^2 + \theta_{14}^2 + \theta_{23}^2 + \theta_{24}^2 + \theta_{34}^2 \right) \right. \nonumber \\
&& \;-\, 27 \left( \theta_{12}\theta_{13}\theta_{23} + \theta_{12}\theta_{14}\theta_{24} + \theta_{13}\theta_{14}\theta_{34} 
+ \theta_{23}\theta_{24}\theta_{34} \right) \nonumber \\
&&\;+\, 27 \left. \left( \theta_{12}\theta_{13}\theta_{24}\theta_{34} + \theta_{12}\theta_{14}\theta_{23}\theta_{34} 
+ \theta_{13}\theta_{14}\theta_{23}\theta_{24} \right) \right].
\label{F_s_indep_Pi}
\ENA
We evaluate $\mathcal{F}^{\text{s-indep}}_{\Pi}$ for specific 
equal-sided trispectrum configurations: collinear, square, rhombus and tetrahedral.
Table \ref{m_config_s_indep_Pi} lists the values of  $\mathcal{F}^{\text{s-indep}}_{\Pi}$ for the 
specific configurations $(\kk_1,\kk_2,\kk_3,\kk_4)$, showing that the greatest contribution to $\left[ \psi_{_{1234}}\right]_{\Pi}$
and therefore to the scalar anisotropic stress trispectrum arises from the collinear 
configuration.
%=========================TABLE===========================
\begin{table}
\caption{The value of the $s$-independent terms  $\mathcal{F}^{\text{s-indep}}_{\Pi}$ in four different equal-sided configurations $(\kk_1,\kk_2,\kk_3,\kk_4)$ 
with $k_1 \sim k_2 \sim k_3 \sim k_4$ for evaluating the magnetic scalar anisotropic stress trispectrum.\\}
\begin{tabular}{ccc} 
\hline \hline \\
\multicolumn{1}{c}{Configuration} &
% \multicolumn{1}{c}{$(\kk_1,\kk_2,\kk_3,\kk_4)$} &
\multicolumn{1}{c}{$\quad (\theta_{12},\theta_{13},\theta_{14},\theta_{23},\theta_{24},\theta_{34})\quad$} &
\multicolumn{1}{c}{$\mathcal{F}^{\text{s-indep}}_{\Pi}$}
 \\ [3 pt]
\hline \hline
%_/_/_/_/_/_/_/_/_/_/_/_/_/_/_/_/_/_/_/_/_/_/_/_/_/_/_/_/_/_/_/_/
\\
collinear &
% $k_1 \sim k_2 \sim k_3 \sim k_4 $  &
$(1,-1,-1,-1,-1,1)$ &
$14$ \\ [5 pt] 
\hline 
%_/_/_/_/_/_/_/_/_/_/_/_/_/_/_/_/_/_/_/_/_/_/_/_/_/_/_/_/_/_/_/_/
\\
square &
% $k_2 \ll k_1, k_3 $ &
$(0,-1,0,0,-1,0)$&
$5$ \\ [5 pt]
\hline
%_/_/_/_/_/_/_/_/_/_/_/_/_/_/_/_/_/_/_/_/_/_/_/_/_/_/_/_/_/_/_/_/
\\
rhombus &
% $k_1 \sim k_2 \sim k_3$ &
$(\frac{1}{2},-1,-\frac{1}{2},-\frac{1}{2},-1,\frac{1}{2})$ &
$2.1875$ \\ [5 pt]
\hline
%_/_/_/_/_/_/_/_/_/_/_/_/_/_/_/_/_/_/_/_/_/_/_/_/_/_/_/_/_/_/_/_/
\\
tetrahedral &
% $k_1 \sim k_3$  &
$(-\frac{1}{2},0,-\frac{1}{2},-\frac{1}{2},0,-\frac{1}{2})$ &
$-2.3125$ \\ [5 pt] 
%_/_/_/_/_/_/_/_/_/_/_/_/_/_/_/_/_/_/_/_/_/_/_/_/_/_/_/_/_/_/_/_/
% collinear &
% $k_2 \ll k_1, k_3 $  &
% &
% \\
% %_/_/_/_/_/_/_/_/_/_/_/_/_/_/_/_/_/_/_/_/_/_/_/_/_/_/_/_/_/_/_/_/
% &
% $\kkk_1 = \kkk_2 = -\kkk_3$  &
% &
% \\ [3 pt]
% \hline
% %_/_/_/_/_/_/_/_/_/_/_/_/_/_/_/_/_/_/_/_/_/_/_/_/_/_/_/_/_/_/_/_/
% \\
% midpoint &
% $k_1 \sim k_2 \sim \frac{k_3}{2}$ &
% $(1,-1,-1)$ &
% $-9$ \\
% %_/_/_/_/_/_/_/_/_/_/_/_/_/_/_/_/_/_/_/_/_/_/_/_/_/_/_/_/_/_/_/_/
% collinear&
% $ \kkk_1 = \kkk_2 = -\kkk_3$ &
% &
% \\ [3 pt]
\hline \hline
\end{tabular}
% \footnotetext[1]{For the squeezed collinear configuration case, $\mathcal
% F_{\Pi_B \Pi_B \Pi_B }^{\text{I}}$ picks up another term $\bar{\mu}^2 \sim 1$}.
\label{m_config_s_indep_Pi}
\end{table}
% ==============================================================
The values for  $\mathcal{F}^{\text{s-indep}}_{\Pi}$ range from $\approx$ -2 to 14. We adopt a value of 10 
as a typical value for the sum of all $s$-independent terms and denote it by $\xi$. 
We get a mode-coupling integral with an integrand that matches the $\left[\psi_{_{1234}}\right]_{\Omega}$ 
for the $\Omega_B$ $s$-independent equal-sided configuration case I Equation (\ref{calI_I})
\EQ
\left[ \psi_{_{1234}}\right]_{\Pi} = \frac{8\, \xi}{(8\pi p_0)^4} \, {\cal I} = \frac{8\,(3^4)\, \xi}{(8\pi\rho_0)^4} \, {\cal I}
\label{psi_Pi_s_indep}
\EN
where
\EQA
{\cal I} &=& \int d^3 s \; M(s) \; M(\vert \kk_1 + \ssvec \vert) \times \nonumber \\ 
  \Big[ &M& (\left| \kk_1 + \kk_3 +\ssvec \right|) \Big( M(\vert \kk_2  -  \ssvec \vert) 
                                                + M(\vert \kk_4  -  \ssvec \vert) \Big) \nonumber \\
+ &M&(\left| \kk_1 + \kk_2 +\ssvec \right|) \Big( M(\vert \kk_3  -  \ssvec \vert)  
                                                + M(\vert \kk_4  -  \ssvec \vert) \Big) \nonumber \\
+ &M&(\left| \kk_1 + \kk_4 +\ssvec \right|) \Big( M(\vert \kk_2  -  \ssvec \vert) 
                                                + M(\vert \kk_3  -  \ssvec \vert) \Big) \! \Big]. 
\label{calI_I_Pi}
\ENA
The integral ${\cal I}$ is evaluated as earlier to yield the four-point correlation of the magnetic scalar anisotropic stress to be
\EQA
\!\left[ \zeta_{_{1234}} \right]_{\Pi} \; &=&  \;\;\delta(\kk_1 + \kk_2 + \kk_3 + \kk_4) \times \nonumber \\
&& \!\!\!\!\!\!\!\!\!\!\!\!\!\!\!\!\!\!\!\!\!\!\!\!\!\!\!\!\!\!\!\!\!\!\!\!\!
{3^4 \, \xi}\,\frac{8 \, (24\pi)\, A^4 \,k_1^{2n+3} k_2^n k_3^n }{(8\pi\rho_0)^4}\left[ \frac{(2^{n/2})(4n+3)-(n+3)}{(4n+3)(n+3)} \right]\!\!, 
\label{zeta_Pi}
\ENA
or simply expressed, in relation to the four-point correlation of energy density, 
\EQ
\left[ \zeta_{_{1234}} \right]_{\Pi} = 3^4 \, \xi \left[ -\zeta_{_{1234}} \right]_{\Omega}.
\label{zeta_Pi_Omega}
\EN

%---------------------------------------------------------------------------------

\section{Magnetic CMB Trispectrum}
\label{s:MagTrispec}

Having calculated the four-point correlations, in Fourier space, of energy density  $\left[ \zeta_{_{1234}} \right]_{\Omega}$ 
and scalar anisotropic stress $\left[ \zeta_{_{1234}} \right]_{\Pi}$, we can now calculate the CMB trispectrum sourced by each.

\subsection{CMB Trispectrum from Magnetic Energy Density}

For the trispectrum sourced by magnetic energy density $\Omega_B$, we insert Eq. (\ref{zeta_Omega}) into Eq. (\ref{trispec}) 
for the trispectrum and following the approach of \cite{OkamotoHu02,KogoKomatsu06}, we decompose our delta function as
$ \delta(\kk_1 + \kk_2 + \kk_3 + \kk_4) = \int d^3 K \delta(\kk_1 + \kk_2 + \KK)\delta(\kk_3 + \kk_4 - \KK)$.
% of four $k$-vectors (sides 
% of a quadrilateral) into an integral of the product of two delta functions each of three $k$-vectors ( 
% two sides and a common diagonal).
We can then write the trispectrum as
\EQA
\!\!&&T^{m_{_1}m_{_2}m_{_3}m_{_4}}_{l_{_1}l_{_2}l_{_3}l_{_4}}
\!= (4\pi)^4 (-1)^{^{\sum_i l_i}} {\cal R}^4\!\!\int 
\frac{d^3 k_1 d^3 k_2 d^3 k_3 d^3 k_4}{(2\pi)^{12}} \nonumber \\
&& \times \; j_{_{l_1}}(k_{_1}D^*) j_{_{l_2}}(k_{_2}D^*) j_{_{l_3}}(k_{_3}D^*) j_{_{l_4}}(k_{_4}D^*) \nonumber \\
&& \times \; Y^*_{l_1m_1}(\hat{\kk}_{_1}) Y^*_{l_2m_2}(\hat{\kk}_{_2}) Y^*_{l_3m_3}(\hat{\kk}_{_3}) Y^*_{l_4m_4}(\hat{\kk}_{_4}) \nonumber \\  
&& \times \left[ \frac{-(192\pi)A^4}{(8\pi \rho_0)^4} k_1^{2n+3} k_2^n k_3^n 
\left\lbrace \frac{(2^{n/2})(4n+3)-(n+3)}{(4n+3)(n+3)} \right\rbrace \right] \nonumber \\
&& \times \int d^3 K \delta(\kk_1 + \kk_2 + \KK)\delta(\kk_3 + \kk_4 - \KK).
\label{trispec_w_zeta}
\ENA
Using the integral form of the delta functions
\EQA
&&\int d^3 K \delta(\kk_1 + \kk_2 + \KK)\delta(\kk_3 + \kk_4 - \KK) = \nonumber \\
&&\!\!\!\!\!\!\int \!\! \frac{d^3 K}{(2\pi)^6} \!\! \int \!\!d^3 r_1^2 \!\! \int \!\!d^3 r_2^2 e^{(\kk_1 + \kk_2 + \KK)\cdot \rr_1} 
e^{(\kk_3 + \kk_4 - \KK)\cdot \rr_2},
\label{K_deltas_exp}
\ENA
and the spherical wave expansion
\EQ
e^{i \kk_{_j} \! \cdot \,\rr} = 4\pi \sum_{l'=0}^{\infty} i^{l'} j_{_{l'}}(k_{_j} r) \sum_{m'=-l'}^{+l'} 
Y^*_{l'm'}({\hat{\kk}}_{_j}) Y_{l'm'}(\hat{\rr}),
\label{sph_wave_exp}
\EN
we perform the integrals over the angular parts of $(\kk_1,\kk_2,\kk_3,\kk_4,\KK)$, with algebra
similar to \cite{FS07,SS09,TSS10,regan10}, to give
\EQA
\!\!&&T^{m_{_1}\!m_{_2}\!m_{_3}\!m_{_4}}_{\,\,\,\,l_{_1}\,\;l_{_2}\,\;l_{_3}\,\;l_{_4}}
% \!= \left[ (-192\pi) \, \frac{2^{10} \, {\cal R}^4}{(2\pi)^8} \right] \left( \frac{A}{(8\pi \rho_0)} \right)^4 \nonumber \\
\!= \left[ (-768) \, \frac{{\cal R}^4}{\pi^7} \right] \left( \frac{A}{(8\pi \rho_0)} \right)^4 \nonumber \\
\!\! && \times \left\lbrace \frac{(2^{n/2})(4n+3)-(n+3)}{(4n+3)(n+3)} \right\rbrace \nonumber \\ 
&& \times \int dr_1 r_1^2 \int dr_2 r_2^2  \int dk_1 k_1^2 k_1^{2n+3} j_{_{l_1}}(k_{_1}D^*) j_{_{l_1}}(k_{_1}r_1)\nonumber \\
&& \times \!\! \int \!\! dk_2 k_2^2 k_2^n j_{_{l_2}}\!(k_{_2}D^*) j_{_{l_2}}\!(k_{_2}r_1) 
    \!\!  \int \!\! dk_3 k_3^2 k_3^n j_{_{l_3}}\!(k_{_3}D^*) j_{_{l_3}}\!(k_{_3}r_2) \nonumber \\
&& \times \int dk_4 k_4^2       j_{_{l_4}}(k_{_4}D^*) j_{_{l_4}}(k_{_4}r_2) \times \sum_{LM} (-1)^{L-M} \nonumber \\
&& \times \int dK     K^2       j_{_{L}}(Kr_1) j_{_{L}}(-Kr_2) \nonumber \\
&& \times \int d\Omega_{\rrr_1} Y_{l_1m_1}(\rrr_1) Y_{l_2m_2}(\rrr_1) Y_{LM}(\rrr_1)  \nonumber \\
&& \times \int d\Omega_{\rrr_2} Y_{l_3m_3}(\rrr_2) Y_{l_4m_4}(\rrr_2) Y_{L\,-M}(\rrr_2).
\label{trispec_after_k_angular}
\ENA
Here the $K$-integral gives $\delta(r_1 - r_2)\left({\pi}/{2 r_1^2}\right)$ using the spherical Bessel 
function closure relation. This delta function enables us to perform the $r_2$-integral trivially, 
then $r_1$ replaces $r_2$ in the arguments of $j_{_{l_3}}$ and $j_{_{l_4}}$. The angular 
$\rrr_1$ and $\rrr_2$-integrals may be expressed as (e.g. Eq. 5.9.1 (5) of \cite{Varshalovich})
\EQA
&& \int d\Omega_{\rrr_1} Y_{l_1m_1}(\rrr_1) Y_{l_2m_2}(\rrr_1) Y_{LM}(\rrr_1) = \nonumber \\
&& \!  \sqrt{\frac{(2l_1+1)(2l_2+1)(2L+1)}{4\pi}} 
\! \! \begin{pmatrix} l_1 & l_2 & L\\ 0 & 0 & 0 \end{pmatrix} 
\! \! \! \begin{pmatrix} l_1 & l_2 & L\\ m_1 & m_2 & M \end{pmatrix} \nonumber \\
&& \equiv h_{l_1 L \, l_2} \begin{pmatrix} l_1 & l_2 & L\\ m_1 & m_2 & M \end{pmatrix},
\label{h_l1_L_l2}
\ENA
where we have defined $h_{l_1 L \, l_2}$ above, in the same convention as \cite{OkamotoHu02,KogoKomatsu06}.
We use the relation 
\EQ
\left( {A}/{8\pi\rho_0} \right)^4 = \left({2}/{3}\right)^4 \left({\pi}/{k_G}\right)^8 
\left({(n+3)}/{k_G^{n+1}}\right)^4 {V_A}^8, 
\label{A_norm_AlfvenVel_8power}
\EN
where the Alfv\'en velocity $V_A$, in the radiation dominated era, is defined as \cite{Jedam98,SB98},
\EQ 
V_{A}=B_{0}/\left( 16\pi \rho _{0}/3 \right)^{1/2}\approx 3.8\times 10^{-4}\, B_{-9},
\label{Alfven_v_value}
\EN
with $B_{-9}\ \equiv (B_{0}/10^{-9}{\rm Gauss})$.
From the definition of the rotationally invariant angle-averaged trispectrum \cite{Hu01}
\EQA
\!\!T^{m_{_1}m_{_2}m_{_3}m_{_4}}_{l_{_1}l_{_2}l_{_3}l_{_4}}
&=& \sum_{LM} (-1)^{-M} \begin{pmatrix} l_1 & l_2 & L\\ m_1 & m_2 & -M \end{pmatrix} \nonumber \\
&& \times \begin{pmatrix} l_3 & l_4 & L\\ m_3 & m_4 & M \end{pmatrix} T^{l_{_1}l_{_2}}_{l_{_3}l_{_4}} (L)
\label{angle_avg_trispec},
\ENA
we separate out the reduced trispectrum $T^{l_{_1}l_{_2}}_{l_{_3}l_{_4}} (L)$ (referred to as the 
angular averaged trispectrum in \cite{Hu01}), from the full trispectrum.
We again use the spherical Bessel function closure relation to perform the $k_4$-integral that yields 
$\delta(r_1-D^*)\left({\pi}/{2r_1^2}\right)$.  This facilitates the $r_1$-integral that results in 
$r_1 \rightarrow D^*$ in the arguments of $j_{_{l_1}}$, $j_{_{l_2}}$ and $j_{_{l_3}}$. 
The $k_1$, $k_2$ and $k_3$-integrals containing a product of a power-law and $ j_{_l}^2$ 
can be evaluated in terms of Gamma functions (e.g. Eq. 6.574.2 of \cite{Gradshteyn6}). 
For a scale-invariant magnetic index $n \to -3$, we get
\EQA
\left[ T^{l_{_1}l_{_2}}_{l_{_3}l_{_4}} (L) \right]_{\Omega}
\simeq && \,-5.8 \times 10^{-29} \left(\frac{n+3}{0.2}\right)^3 \left(\frac{B_{-9}}{3}\right)^8 \nonumber \\
 && \times \frac{ h_{l_1 L \, l_2} \, h_{l_3 L \, l_4}}{l_1(l_1+1) l_2(l_2+1) l_3(l_3+1)}.
\label{trispec_energy_density}
\ENA
This equation gives us the amplitude of the magnetic CMB trispectrum sourced by the 
energy density $\Omega_B$ of a primordial magnetic field, where we have used ${\mathcal R} \sim -0.04$ \cite{bonvin10}. 
A factor of $1/(D^* k_G)^{4(n+3)}$ also 
appears here and it approaches unity for the case $n \to -3$ (a scale-invariant magnetic field index). When we evaluate the
magnetic trispectrum for a near scale-invariant index $n=-2.8$, this factor has a value 
$\sim 1/1500$. It then turns out that this factor is almost entirely canceled by the simultaneous increase in the value of the $k$-integrals
when evaluated for $n=-2.8$ rather than $n=-3$. 

For the case II - collinear configuration case, proceeding from Eq. (\ref{zeta_coll}) in exactly the same way as case I, 
we find that the amplitude of the collinear configuration trispectrum is 
\EQA
\left[ T^{l_{_1}l_{_2}}_{l_{_3}l_{_4}} (L) \right]_{\Omega}
\simeq && \,3.9 \times 10^{-29} \left(\frac{n+3}{0.2}\right)^3 \left(\frac{B_{-9}}{3}\right)^8 \nonumber \\
 && \times \frac{ h_{l_1 L \, l_2} \, h_{l_3 L \, l_4}}{l_1(l_1+1) l_2(l_2+1) l_3(l_3+1)},
\label{trispec_energy_density_collinear}
\ENA
which is similar in magnitude to the case I trispectrum, but of positive sign.

\subsection{CMB Trispectrum from Magnetic Scalar Anisotropic Stress}

The scalar anisotropic stress trispectrum $\left[ T^{l_{_1}l_{_2}}_{l_{_3}l_{_4}} (L)\right]_{\Pi}$ can be calculated 
in an analogous manner to the calculation presented above for case I $s$-independent $\left[ T^{l_{_1}l_{_2}}_{l_{_3}l_{_4}} (L)\right]_{\Omega}$.
Using Equations (\ref{trispec_Pi}) and (\ref{zeta_Pi}) we obtain
\EQA
\left[ T^{l_{_1}l_{_2}}_{l_{_3}l_{_4}} (L)\right]_{\Pi}
&\simeq&  \left( 3~\frac{\mathcal{R}_p}{\mathcal{R}} \right)^4 ~\xi~\left[- T^{l_{_1}l_{_2}}_{l_{_3}l_{_4}} (L)\right]_{\Omega} \nonumber \\
&\simeq& 1.1 \times 10^{-19}  ~\left( \frac{\xi}{10} \right) \left(\frac{n+3}{0.2}\right)^3 \left(\frac{B_{-9}}{3}\right)^8 \nonumber \\
&& \quad \times \frac{ h_{l_1 L \, l_2} \, h_{l_3 L \, l_4}}{l_1(l_1+1) l_2(l_2+1) l_3(l_3+1)}.
\label{trispec_aniso_stress_result}
\ENA

We see that the amplitude of the trispectrum sourced by $\Pi_B$ for equal-sided quadrilateral configurations is 
approximately $10^{10}$ times larger than that sourced by $\Omega_B$. 
Here, we have used $T_B \simeq 10^{14}$ GeV (corresponding to the reheating temperature) and $T_{\nu} \simeq 10^{-3}$ GeV.

%++++++++++++++++++++++++++++++++++++++++++++++++++++++++++++++++++++

\section{Flat-Sky Calculation of Scalar Anisotropic Stress CMB Trispectrum}
\label{s:FlatSky}

We now consider a flat-sky analysis of the trispectrum.
The flat-sky limit allows us to avoid the approximate treatment of 
the angular terms involving $\kkk_i$
while performing the $\kkk$ angular integrals that led to Equation (\ref{trispec_after_k_angular}). Therefore, to get a more accurate estimate of 
the $s$-independent anisotropic stress trispectrum, we now adopt the flat-sky limit for the CMB temperature anisotropy and recompute the trispectrum.

In the flat-sky limit \cite{zald_seljak_97,hu2000,boubekeur09}, the CMB temperature fluctuations on the sky are expanded in terms of plane waves using a Fourier basis 
rather than a spherical harmonic basis,
\EQA
\frac{\Delta T}{T}(\nn) &=&  \int \frac{d^2 l}{\left( 2 \pi \right)^2} a_{_{\ell}} e^{i \, \llvec \cdot \nn}, \nonumber \\
a_{_{\ell}} &=& \int d^2 n \frac{\Delta T}{T}(\nn) e^{-i \, \llvec \cdot \nn}.
\label{flat_sky_defn}
\ENA
In the flat-sky co-ordinates, $\ell = (\ell_x,\ell_y)$ is a two-dimensional vector on the plane of the sky and $n_z$ is a constant equal 
to unity at linear order. In order to check the validity of our flat-sky technique, 
we first 
computed the magnetic energy density bispectrum. We find a value 
for the flat-sky bispectrum of order $\approx 10^{-23}$, which agrees well 
with the original full-sky result \cite{SS09}.
This encourages us to proceed to the flat-sky limit calculation of the scalar anisotropic stress trispectrum.

The magnetic Sachs-Wolfe effect for scalar anisotropic stress is given by
\EQA
\frac{\Delta T}{T}(\nn) &=&  \mathcal{R}_p ~ \Pi_B(\xx_0 -\nn D^*) \\ \nonumber
&=& \int \frac{d^3 k}{\left( 2 \pi \right)^3} \mathcal{R}_p \, \Pi_B(\kk) e^{i \, \kk \cdot \left( \xx_0 -\nn D^*\right)} \nonumber \\
&=& \mathcal{R}_p \int \frac{d^3 k}{\left( 2 \pi \right)^3} \, \Pi_B(\kk) e^{-i (\kk \cdot \nn) D^*}.
\label{flat_PiB}
\ENA
where in the last line we set the observer's position $\xx_0$ to the origin. 

The flat-sky limit is accurate for $\ell \gtrsim 40 $ \cite{zald_seljak_97,hu2000,boubekeur09} whereas the 
Sachs-Wolfe contribution is appreciable for $\ell \lesssim 100$ (but dominant only till $\ell \lesssim 50 $) \cite{lewis_full_bispec}. 
Therefore, there exists an appreciable range of overlap $40 \lesssim \ell \lesssim 100$ in harmonic space, 
where we can treat the Sachs-Wolfe contribution to the CMB temperature anisotropy in the flat-sky limit. 
 
In the flat-sky limit, $n_z$ is constant and is unity to linear order hence $\nn \cdot \kk  \rightarrow \mm \cdot \kk_{\perp} + k_z $ which gives
\EQA
a_{_{\ell}} &=& \int d^2 n \left( \frac{\Delta T}{T}(\nn) \right)_{\text{\!\!flat sky}} \, e^{-i \, \llvec \cdot \nn} \\
&=& \mathcal{R}_p \int \frac{d^3 k}{\left( 2 \pi \right)^3} \, \Pi_B(\kk) \, e^{-i \, k_z D^*} \int d^2 m \,
e^{-i \, \mm \cdot \left( \llvec + \kk_{\perp} D^* \right)}. \nonumber
\ENA
The $m$-integral gives a delta function for $\kk_{\perp}$
\EQA
\int d^2 \! &m& e^{-i \, \mm \cdot \left( \llvec + \kk_{\perp} D^* \right)} = (2 \pi)^2 \delta^{(2)}\left( \llvec + \kk_{\perp} D^* \right) \nonumber \\
&=& \left( \frac{2 \pi}{D^*} \right)^2 \delta^{(2)}\left( \frac{\llvec}{D^*} + \kk_{\perp} \right)
\ENA
to yield
\EQ
a_{_{\ell}} = \frac{\mathcal{R}_p}{(D^*)^2} \int_{\infty}^{\infty} \frac{d k_z}{2 \pi}  \, \Pi_B \!
\! \left( \kk_{\perp}\!=\!\frac{-\llvec}{D^*}, \, k_z \right) \, e^{-i \, k_z D^*}.
\EN

This flat-sky $a_{_{\ell}}$ for magnetic scalar anisotropic stress can then be used to calculate the 
corresponding trispectrum in the flat-sky limit
\EQ
\langle a_{\ell_1}a_{\ell_2}a_{\ell_3}a_{\ell_4} \rangle = \left(\frac{\mathcal{R}_p}{2\pi}\right)^4
\left[ \prod_{i=1}^4 \int_{\infty}^{\infty} \! d k_{i_z} \frac{e^{-i \,k_{i_z} D_i^*}}{{(D_i^*)^2}} \right] 
\!\zeta^{\text{fs}}_{_{1234}}
\label{flatsky_4pt_al}
\EN
where $\zeta^{\text{fs}}_{_{1234}}$ is the four-point correlation of magnetic scalar anisotropic stress 
in the flat-sky limit
\EQ
\zeta^{\text{fs}}_{_{1234}} = \left\langle \left[ \prod_{i=1}^4 \, \Pi_B \! \! 
\left( \! \kk_{i_{\perp}}\!\!=\!\frac{-\llvec_i}{D_i^*}, \, k_{i_z} \!\! \right) \right] \right\rangle.
\EN 
As before in the full-sky for $\zeta_{_{1234}}$ (Eqs.~\ref{zeta_Pi_definition}, \ref{zeta_Pi}), a four-point correlation of $\Pi_B$ produces
delta functions times a mode coupling integral $\psi$. 
\EQA
\zeta^{\text{fs}}_{_{1234}}  = & \delta &\!\!\left( \! k_{1_z} \!\! + \! k_{2_z}\!\! + \!k_{3_z}\!\!+\!k_{4_z} \!\right)\! \\ \nonumber
&\times& \delta^{(2)} \!\! \left( \!\frac{\llvec_1}{D_1^*}\!\!+\!\frac{\llvec_2}{D_2^*}\!\!+\!\frac{\llvec_3}{D_3^*}\!\!+
\!\frac{\llvec_4}{D_4^*} \! \right) \left[ \psi_{_{1234}}\right]^{\text{fs}}_{\Pi}
\ENA
If we take the $D_i^*$'s  to be similar, we find
\EQA
\zeta^{\text{fs}}_{_{1234}}  = & \delta &\!\!\left( \! k_{1_z} \!\! + \! k_{2_z}\!\! + \!k_{3_z}\!\!+\!k_{4_z} \!\right)\! \\ \nonumber
&\times& (D^*)^2 \,\, \delta^{(2)} \!\! \left( \! \llvec_1\!\!+\!\llvec_2\!\!+\!\llvec_3\!\!+\!\llvec_4 \! \right)
\left[ \psi_{_{1234}}\right]^{\text{fs}}_{\Pi}
\ENA
Here the mode-coupling integral $\psi$ is
\EQ
\left[ \psi_{_{1234}}\right]^{\text{fs}}_{\Pi} = \frac{8\,\mathcal{F}^{\text{s-indep}}_{\Pi}}{(8\pi p_0)^4} \, {\cal I}
\label{psi_flatsky_Pi_s_indep}
\EN
where the integral $\mathcal{I}$ is the same as the one given by Eq. (\ref{calI_I_Pi}) and the 
$s$-independent angular terms for $\Pi_B$ are denoted by $\mathcal{F}^{\text{s-indep}}_{\Pi}$ 
given by Eq. (\ref{F_s_indep_Pi}). In the flat-sky approach we perform the mode-coupling integral for general
values of $\kkk_i \cdot \kkk_j$ and later evaluate the trispectrum for particular configurations that are not necessarily 
equal-sided. The first term (out of six terms) of integral $\mathcal{I}$ is
\EQ
{\cal I}^{\text{fs}}_{\rm (1)} \!\simeq \!4 \pi \!A^4 k_1^{2n+3} k_2^n 
\left[ \!\frac{\left( k_1^2 + 2 k_1 k_3 \theta_{13} + k_3^2 \right)^{n/2}}{n+3} - \frac{k_3^n}{4n+3}\! \right]
\label{flat_Ii}
\EN
Whereas, in the full-sky $\Pi_B$ calculation we chose a representative value $\xi$ for 
$\mathcal{F}^{\text{s-indep}}_{\Pi}$, we now integrate over all 14 terms of 
$\mathcal{F}^{\text{s-indep}}_{\Pi}$ in the $k_{i_z}$ integrals.

\begin{figure}[tbp]
\centering
\epsfig{file=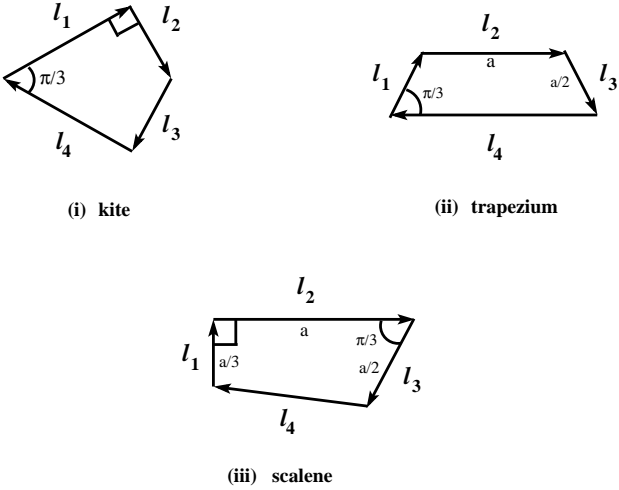,width=0.95\linewidth,clip=}
\caption{The three specific $\ell$ wavevector configurations (i) kite, (ii) trapezium (both cyclic quadrilaterals) 
and (iii) scalene (an irregular convex quadrilateral)
used to evaluate the flat-sky magnetic scalar anisotropic trispectrum. Trispectrum configuration shapes (i) and (ii) are also discussed in 
\cite{hindmarsh,lewis_shape}}
\label{fig_3flatsky_configs}
\end{figure}

For each of the 6 terms of $\mathcal{I}$, the delta function of $k_{i_z}$ is used to perform that 
particular $k_{i_z}$ integral (one out of four) for which the variable $k_i$ that does not appear in the arguments 
of the magnetic spectrum $M$. 
This introduces substitutions in the angular structure $\mathcal{F}^{\text{s-indep}}_{\Pi}$. 
Then the remaining three $k_{i_z}$ integrals are performed numerically and evaluated for several types of configurations. 
We use the relation for the flat-sky trispectrum (connected part) \cite{OkamotoHu02,KogoKomatsu06} 
\EQ
\langle a_{_{\ell_1}} a_{_{\ell_2}} a_{_{\ell_3}} a_{_{\ell_4}} \rangle = (2\pi)^2 
\delta^{(2)} \!\! \left( \! \llvec_1\!\!+\!\llvec_2\!\!+\!\llvec_3\!\!+\!\llvec_4 \! \right) 
T^{(\ell_1,\ell_2)}_{(\ell_3,\ell_4)} (L)
\EN
to get $\left[ T^{(\ell_1,\ell_2)}_{(\ell_3,\ell_4)} (L)\right]_{\Pi}$ from the four-point correlation of $a_{_{\ell}}$.
The product of the mode coupling integral $\mathcal{I}$ and the three $k_{i_z}$ integrals is denoted by $\sigma$.
Table (\ref{sigma_flatsky_config_s_indep_Pi}) shows different values of $\sigma$ for different $\ell$-space configurations 
with parameters $q_{ab} = l_{_{_a}} / l_{_{b}}$ (ratio of different sides) and $\lllvec_i \cdot \lllvec_j$ 
(cosine of the angle between sides). 
We note that all the configurations thus evaluated in the flat-sky approach (for all s-independent terms) give 
a negative $\sigma$ that lead to a negative value of the trispectrum. 
% \textbf{\textit{(***How/Where to mention configurations that blow up- collinear, square, rectangle)}}
\!\!\footnote[1]{For some highly symmetrical configurations which have two $\llvec$ vectors exactly anti-parallel and 
of equal magnitude, the $k_z$ integral becomes singular in the flat-sky limit. However, this is due to the exact 
$\kk_{i_{\perp}}\!\!=\!\frac{-\llvec_i}{D_i^*}$ map which is enforced in this limit. If this were relaxed then we expect 
this mathematical pathology to be just an integrable singularity. The measure of such configurations in $d^3k$ is 
expected to go to zero faster than the reciprocal of the integrand.}
\EQA
\left[ T^{(\ell_1,\ell_2)}_{(\ell_3,\ell_4)} (L)\right]_{\Pi}
&\simeq& 3.94 \times 10^{-19}  ~\left( \frac{\sigma}{10} \right) \left(\frac{n+3}{0.2}\right)^3 \left(\frac{B_{-9}}{3}\right)^8 \nonumber \\
&& \quad \times \frac{1}{l_1^2 l_2^2 l_3^2}.
\label{trispec_flatsky_aniso_stress}
\ENA
We see that the flat-sky evaluation of the scalar anisotropic stress trispectrum with $s$-independent terms 
results in trispectra that are negative and roughly an order of magnitude larger in absolute magnitude 
than the corresponding full-sky trispectrum with $s$-independent terms (with $\xi \approx 10$).
The flat-sky and full-sky trispectra are related by
\EQ
T^{(\ell_1,\ell_2)}_{(\ell_3,\ell_4)} (L) h_{l_1 L \, l_2} \, h_{l_3 L \, l_4} 
\approx T^{l_{_1}l_{_2}}_{l_{_3}l_{_4}} (L).
\EN
This allows us to compare the flat-sky trispectrum directly to the full-sky trispectrum form 
given in Eq.(\ref{trispec_aniso_stress_result})

\begin{widetext}

%=========================TABLE===========================
\begin{table}
\caption{The value of $\sigma$ [the product of the integral $\mathcal{I}$ Eq.(\ref{calI_I_Pi}) 
and $k_{i_z}$ integrals Eq.(\ref{flatsky_4pt_al})]
% over the $s$-independent terms $\mathcal{F}^{\text{s-indep}}_{\Pi}$ 
for the three different trispectrum configurations 
% $(\ell_1,\ell_2,\ell_3,\ell_4)$
(shown in Fig. \ref{fig_3flatsky_configs})
considered for the flat-sky magnetic scalar anisotropic stress trispectrum (Eq. \ref{trispec_flatsky_aniso_stress}).\\}

\begin{tabular}{cccc} 
\hline \hline \\ 
\multicolumn{1}{c}{Configuration} &
% \multicolumn{1}{c}{$(\kk_1,\kk_2,\kk_3,\kk_4)$} &
\multicolumn{1}{c}{$\quad (q_{12},q_{13},q_{14},q_{23},q_{24},q_{34})\quad$} &
\multicolumn{1}{c}{$\quad (\lllvec_1 \! \cdot \! \lllvec_2, \,\lllvec_1 \! \cdot \! \lllvec_3, \,
\lllvec_1 \! \cdot \! \lllvec_4, \,\lllvec_2 \! \cdot \! \lllvec_3, \,\lllvec_2 \! \cdot \! \lllvec_4, \,\lllvec_3 \! \cdot \! \lllvec_4)\quad$} &
\multicolumn{1}{c}{$\sigma$}
 \\ [5 pt]
\hline \hline
%_/_/_/_/_/_/_/_/_/_/_/_/_/_/_/_/_/_/_/_/_/_/_/_/_/_/_/_/_/_/_/_/
\\
kite &
% $k_1 \sim k_2 \sim k_3$ &
$(\sqrt{3},\sqrt{3},1,1,1/\sqrt{3},1/\sqrt{3})$&
$(0,-\sqrt{3}/2,-1/2,1/2,-\sqrt{3}/2,0)$ &
$-15.2$ \\ [5 pt]
\hline
%_/_/_/_/_/_/_/_/_/_/_/_/_/_/_/_/_/_/_/_/_/_/_/_/_/_/_/_/_/_/_/_/
\\
trapezium &
% $k_2 \ll k_1, k_3 $ &
$(2,2/3,2,1/3,1,3)$&
$(1/2,-1,1/2,-1/2,-1/2,-1/2)$ &
$-84.6$ \\ [5 pt]
\hline
%_/_/_/_/_/_/_/_/_/_/_/_/_/_/_/_/_/_/_/_/_/_/_/_/_/_/_/_/_/_/_/_/
\\
scalene &
% $k_1 \sim k_3$  &
$(1/3,2/3,0.4406,2,1.322,0.6609)$ &
$(0,-\sqrt{3}/2,0.1317,-1/2,-0.9912,0.3815)$ &
$-14.2$ \\ [5 pt] 
\hline \hline
\end{tabular}
% \footnotetext[1]{For the squeezed collinear configuration case, $\mathcal
% F_{\Pi_B \Pi_B \Pi_B }^{\text{I}}$ picks up another term $\bar{\mu}^2 \sim 1$}.
\label{sigma_flatsky_config_s_indep_Pi}
\end{table}
% ==============================================================

\end{widetext}

%++++++++++++++++++++++++++++++++++++++++++++++++++++++++++++++++++++++++++++++++++

\section{Primordial Magnetic Field Constraints}
\label{s:BLimits}

We can now compare our magnetic trispectra with the 
Sachs-Wolfe contribution to the standard CMB trispectrum 
sourced by non-linear terms in the inflationary perturbations 
calculated by Okamoto \& Hu \cite{OkamotoHu02} and Kogo \& Komatsu \cite{KogoKomatsu06} (also see \cite{ByrnesSasakiWands06}).
\EQA
T^{l_{_1}l_{_2}}_{l_{_3}l_{_4}} (L) \approx &9& \, C_{l_2}^{SW} C_{l_4}^{SW} \left[ \left( 25/9 \right) \, \tau_{NL} C_{L}^{SW} \right. \nonumber \\
 &+& \left. 6 \, g_{NL} \left( C_{l_1}^{SW} + C_{l_3}^{SW} \right) \right] \, h_{l_1 L \, l_2} \, h_{l_3 L \, l_4}
\ENA
We neglect the $g_{NL}$ term that places far weaker constraints on the trispectrum compared to the $\tau_{NL}$ term considering the 
the current limits on $g_{NL}$ from WMAP \cite{smidt10} and current limits on $\tau_{NL}$ from \textit{Planck} \cite{Planck_NG}. 
The CMB angular power spectrum $C_l^{SW}$ in the Sachs-Wolfe approximation for a scale-invariant primordial 
power spectrum for $\Phi$ is
\EQ
C_l^{SW} = \frac{2}{9\,\pi} \int k^2 dk P_{\Phi}(k) j_{_{l}}^2(kr_*) = \frac{A_{\Phi}}{l(l+1)},
\EN
where $A_{\Phi}$ is the amplitude of scalar potential perturbations. This gives
\EQA
\!\!\!\!\!\!\!\!\! T^{l_{_1}l_{_2}}_{l_{_3}l_{_4}} (L) 
&\approx& 25 \, C_{l_2}^{SW} \,C_{l_4}^{SW} \,C_{L}^{SW} \,\tau_{NL} \, h_{l_1 L \, l_2} \, h_{l_3 L \, l_4} \nonumber \\
&\approx& 25 \,A_{\Phi}^3 \,\tau_{NL} \,\frac{ h_{l_1 L \, l_2} \, h_{l_3 L \, l_4}}{l_2(l_2+1) l_4(l_4+1) L(L+1)} \nonumber \\
&\approx& 25 \,A_{\Phi}^3 \,\tau_{NL} \,\frac{ h_{l_1 L \, l_2} \, h_{l_3 L \, l_4}}{l_1(l_1+1) l_2(l_2+1) l_3(l_3+1)} \nonumber \\
&& \quad \times \frac{{l_1(l_1+1) l_3(l_3+1)}}{{l_4(l_4+1) L(L+1)}}\nonumber \\
&\approx& 25 \, A_{\Phi}^3 \, \tau_{NL} \, \frac{ h_{l_1 L \, l_2} \, h_{l_3 L \, l_4}}{l_1(l_1+1) l_2(l_2+1) l_3(l_3+1)} \, q,
\ENA 
where we also define a factor 
$q = \!\left[{l_1(l_1+1) l_3(l_3+1)}\right]\!/\!\left[{l_4(l_4+1) L(L+1)} \right]$
which is of order unity for many configurations. 
To calculate the value of $A_{\Phi}$ we begin with the most recent \textit{Planck} 2013 data release value for the amplitude of 
scalar curvature perturbations on \cite{Planck_Parameters} $A_s = 2.2 \times 10^{-9}$ at a pivot scale  $k_0 = 0.05 \text{ Mpc}^{-1}$.
For the purpose of the Sachs-Wolfe contribution we then calculate the scalar amplitude at the larger scale of $k_0 = 0.002 \text{ Mpc}^{-1}$ 
using the \textit{Planck} 2013 value for the scalar spectral index $n_s = 0.96$. After converting from curvature to potential we get 
$A_{\Phi} = 6.96 \times 10^{-10}$.
Hence, we find the amplitude for the Sachs-Wolfe contribution to the 
standard CMB trispectrum sourced by inflationary perturbations to be  
\EQ 
T^{l_{_1}l_{_2}}_{l_{_3}l_{_4}} (L) 
\approx 8.4 \times 10^{-27} \tau_{NL} \frac{ h_{l_1 L \, l_2} \, h_{l_3 L \, l_4}}{l_1(l_1+1) l_2(l_2+1) l_3(l_3+1)} \, q. 
\label{trispectrum_tauNL}
\EN
Equation (\ref{trispectrum_tauNL}) is of the same form as Eq. (\ref{trispec_energy_density}) 
and Eq. (\ref{trispec_aniso_stress_result}) for the magnetic field-induced trispectra, facilitating direct comparison of trispectra values.

\subsection{Limits from Magnetic Energy Density - Case I}

We can put upper limits on the primordial magnetic field by comparing the magnetic energy density trispectrum Eq. (\ref{trispec_energy_density}) 
with the inflationary trispectrum Eq. (\ref{trispectrum_tauNL}), although stronger constraints follow from magnetic anisotropic stress.
We take the two-sigma upper limit value on $\tau_{NL}$ reported in the \textit{Planck} 2013 data release: $\tau_{NL} < 2,800$ \cite{Planck_NG} 
and use it also as a lower limit for possible negative values of $\tau_{NL}$ i.e. $\left| \tau_{NL} \right| < 2,800$. 
This is tighter than the $\tau_{NL} > -6,000$ negative-sided limit from WMAP5 data \cite{smidt10} that we employed in \cite{TSS11_Trispec_Letter}.
Magnetic field limits are obtained by taking the one-eighth power of the appropriate ratio of trispectra, 
which gives $B_0 \lesssim 19 \text{ nG}$
at a scale of $k_G = 1 h$ Mpc$^{-1}$ for a magnetic spectral index of $n=-2.8$. 
This trispectrum limit is almost a factor of 2 stronger than the 
bispectrum upper limit $B_0 \lesssim 35 \text{ nG}$ found for
magnetic energy density \cite{SS09} for the same scale and magnetic index.

We note that if we update the value of ${\cal R}$ used in the earlier bispectrum calculation \cite{SS09} to the currently adopted value of
${\cal R}$ \cite{bonvin10} then the magnetic energy density bispectrum yields a 
tighter upper limit of $B_0 \lesssim 30 \text{ nG}$. The trispectrum constraint we calculated above, $B_0 \lesssim 19 \text{ nG}$, 
seems significantly stronger than the bisectrum constraint (by a factor of 1.6). However, since the energy density bispectrum 
calculation \cite{SS09} was performed, the $f_{NL}^{loc}$ two-sigma upper limit has tightened from 
$\approx 100$ (WMAP5) \cite{WMAP5} to 74 (WMAP7) \cite{komatsu_wmap7} to 14.3 (\textit{Planck} 2013) \cite{Planck_NG}. 
Recalculation of the magnetic field constraint from the magnetic energy density bispectrum, now using $f_{NL}^{loc} < 14.3$, 
yields $B_0 \lesssim 22 \text{ nG}$. We see that the corresponding magnetic energy density trispectrum limit (19 nG) 
found in this work is, nevertheless, slightly stronger than the updated bispectrum limit.

\subsection{Limits from Magnetic Energy Density - Collinear Configuration}

We have also calculated the magnetic energy density trispectrum considering all the angular terms that appear for the collinear configuration (case II).
Comparing the collinear configuration energy density trispectrum Eq.(\ref{trispec_energy_density_collinear}) to the inflationary trispectrum 
Eq.(\ref{trispectrum_tauNL}) leads to upper limits on the primordial magnetic field of $B_0 \lesssim 20 \text{ nG}$,
having employed the positive-sided limit $\tau_{NL} < 2,800$ \cite{Planck_NG}. 
This $B_0$ limit from the collinear configuration trispectrum that considers the full mode-coupling integral over all angular terms
is similar to the limit above from case I: only $\ssvec$-independent angular terms for any equal-sided configuration.

\subsection{Limits from Scalar Anisotropic Stress}

The trispectrum from magnetic scalar anisotropic stress Eq.(\ref{trispec_aniso_stress_result}) was found to be $10^{10}$ times
larger than the trispectrum from magnetic energy density. Comparing it with the trispectrum from inflationary perturbations (Eq.\ref{trispectrum_tauNL}) 
gives a much stronger magnetic field constraint of
\EQ
B_0 \lesssim 0.9 \text{ nG},
\label{B_Pi_tau}
\EN
using the positive-sided limit $\tau_{NL} < 2,800$ from the \textit{Planck} 2013 data release \cite{Planck_NG}. 

This $B_0 \lesssim 0.9 \text{ nG}$ limit is over two and a half times as strong as the $B_0$ limit (2.4 nG) 
obtained from the $\Pi_B$ bispectrum \cite{TSS10}.
In addition, for those theories of inflation, which lead to $\tau_{NL} = \left( 6/5 \,f_{NL}\right)^2 $
we could perhaps use the 
relatively tighter limits on $f_{NL}$.  
The two-sigma limits on $f_{NL}^{loc}$ are $-8.9 < f^{loc}_{NL} < 14.3$, obtained from searching for the CMB primordial bispectrum signal 
in \textit{Planck} 2013 data \cite{Planck_NG}. This gives a primordial magnetic field limit of 
\EQ
B_0 \lesssim 0.7 \text{ nG},
\label{B_Pi_fNL}
\EN
for both the negative and positive $f_{NL}^{loc}$ limits separately. We employ the local configuration $f_{NL}$ limits as the 
uncertainties $\sigma_{f_{NL}}$ in the other orthogonal and equilateral configurations are about an order of magnitude larger.

\subsection{Limits from Scalar Anisotropic Stress - Flat-Sky}

We can also compare the flat-sky calculation of the scalar anisotropic stress trispectrum to the 
trispectrum from inflationary perturbations (Eq. \ref{trispectrum_tauNL})
and obtain magnetic field limits using the negative-sided limit of $\left| \tau_{NL} \right| < 2,800$ to get
\EQ
B_0 \lesssim 0.6-0.8 \text{ nG}.
\label{B_Pi_flatsky_tau}
\EN
The range of magnetic field upper limits reflects the range of $\sigma$ values (-84.6 to -14.2) 
in Table (\ref{sigma_flatsky_config_s_indep_Pi}) for different configurations of the 
flat-sky trispectrum. As before, we may again consider those theories of inflation 
which lead to $\tau_{NL} = \left( 6/5 \,f_{NL}\right)^2 $ and use the relatively tighter limits on $f_{NL}$, i.e 
$-8.9 < f^{loc}_{NL} < 14.3$ \cite{Planck_NG} to place magnetic field upper limits of
\EQ
B_0 \lesssim 0.4-0.6 \text{ nG},
\label{B_Pi_flatsky_fNL}
\EN
where we take the combined effect of the slightly different (positive and negative) limits for $f_{NL}$ 
as well as the range of values of $\sigma$ 
to arrive at the range of $B_0$ upper limits. 

For magnetic scalar anisotropic stress, the flat-sky trispectra values give magnetic field upper limits 
that are slightly stronger but consistent with the sub-nanoGauss values derived from the full-sky trispectrum.

\subsection{Limits from Inflationary Magnetic Curvature Mode}

Recently, Bonvin et al. \cite{bonvin13,bonvin12} have found a magnetic mode in the curvature perturbation 
that is present only when magnetic fields are generated at inflation. This magnetic mode 
is always scale-invariant and is absent when magnetogenesis occurs causally e.g. via a phase transition. 
This inflationary magnetic mode is seen to exist in addition to the compensated and passive modes 
and dominates over them in the CMB anisotropy. The ratio of the passive mode power spectrum to the
new inflationary magnetic mode power spectrum is proportional to $\epsilon^2$ where $\epsilon \sim 10^{-2}$ is the
inflationary slow-roll parameter. We calculate the passive to inflationary power spectrum ratio using the relation
given between Equations (45) and (46) in Bonvin et al. \cite{bonvin13}, for $n \rightarrow -3$,
\EQA
\frac{C_l^{\text{passive}}}{C_l^{\text{infl. mag.}}} \simeq 
&\epsilon^2& \, \ln^2 \!\left( \frac{\eta_*}{\eta_{\nu}}\right) \, \left( \frac{\eta_*}{\eta_0} \right)^{2n+6} \,
\frac{\Gamma\left(-n-2\right)}{\Gamma \left(-n-\frac{3}{2}\right)} \nonumber \\
&\times& l^{2n+6} \, \ln^2 \left( \frac{\eta_*}{\eta_0}\right),
\ENA
to find 
\EQ
\frac{C_l^{\text{passive}}}{C_l^{\text{infl. mag.}}} \simeq 4.7 \times 10^{-5}.
\EN
Now consider the magnetic CMB trispectrum sourced by this inflationary magnetic mode. We assume the trispectra ratio 
scales approximately as the power spectrum ratio squared and magnetic field constraint will come from one-eighth 
power of trispectra ratio. The magnetic field constraint is then found to be significantly stronger than from magnetic passive modes 
(i.e. scalar anisotropic stress $\Pi_B$) roughly by a factor $\approx (4.7 \times 10^{-5})^{-0.25} \approx 12$. The
magnetic field upper limit from the inflationary magnetic mode CMB trispectrum is then
\EQ
B_0 \lesssim 0.05 \text{ nG} \qquad \text{i.e } \qquad B_0 \lesssim 50 \text{ picoGauss}.
\label{B_Infl_Mag}
\EN

For this inflationary magnetic mode, the trispectrum, as well as other CMB correlations, give magnetic field upper limits 
that are an order of magnitude stronger than those derived from the magnetic passive mode (scalar anisotropic stress) alone.
Clearly, the new inflationary magnetic mode presented by Bonvin et al. \cite{bonvin13}  seems to place stronger constraints on 
primordial magnetic fields from its CMB correlations and we hope to return to this in greater detail in future work.

\begin{widetext}

%=========================TABLE===========================
\begin{table}
\caption{Comparison of upper limits on primordial magnetic fields from magnetic mode contributions to the 
CMB power spectra, bispectra and trispectra (this work). We quote limits derived for close to scale-invariant magnetic fields
and an early generation epoch ($10^{14}$ GeV) for magnetic passive modes.\\}
\begin{tabular}{cccc} 
\hline \hline \\
\multicolumn{1}{c}{\,\,\,\,\, CMB Probe \,\,\,\,\,} &
% \multicolumn{1}{c}{$(\kk_1,\kk_2,\kk_3,\kk_4)$} &
\multicolumn{1}{c}{\,\,\,Magnetic modes\,\,\,} &
\multicolumn{1}{c}{Magnetic field upper limit $B_0$ (nG)} &
\multicolumn{1}{c}{\,\,\,Reference\,\,\,}
 \\ [3 pt]
\hline \hline
%_/_/_/_/_/_/_/_/_/_/_/_/_/_/_/_/_/_/_/_/_/_/_/_/_/_/_/_/_/_/_/_/
\\
Power Spectrum &
scalar, vector \&  tensor &
3.4 &
\cite{Planck_Parameters} \\ [5 pt] 
\hline  
%_/_/_/_/_/_/_/_/_/_/_/_/_/_/_/_/_/_/_/_/_/_/_/_/_/_/_/_/_/_/_/_/
\\
Bispectrum &
energy density&
\,\,22 \!\footnote{The magnetic field upper limit from \cite{SS09} has been updated 
with the current values for $\mathcal{R}$ and current upper limit for $f_{NL}$} &
\cite{SS09} \\ [5 pt]
%\hline
%_/_/_/_/_/_/_/_/_/_/_/_/_/_/_/_/_/_/_/_/_/_/_/_/_/_/_/_/_/_/_/_/
\\
Bispectrum &
scalar anisotropic stress &
2.4 &
\cite{TSS10} \\ [5 pt]
%\hline
%_/_/_/_/_/_/_/_/_/_/_/_/_/_/_/_/_/_/_/_/_/_/_/_/_/_/_/_/_/_/_/_/
\\
Bispectrum &
vector &
10 &
\cite{Shiraishi_vector} \\ [5 pt]
%\hline
%_/_/_/_/_/_/_/_/_/_/_/_/_/_/_/_/_/_/_/_/_/_/_/_/_/_/_/_/_/_/_/_/
\\
Bispectrum &
tensor &
3.2 &
\cite{Shiraishi_tensor_WMAP7} \\ [5 pt]
\hline
%_/_/_/_/_/_/_/_/_/_/_/_/_/_/_/_/_/_/_/_/_/_/_/_/_/_/_/_/_/_/_/_/
\\
Trispectrum &
energy density &
19 &
this work \\ [5 pt]
%\hline
%_/_/_/_/_/_/_/_/_/_/_/_/_/_/_/_/_/_/_/_/_/_/_/_/_/_/_/_/_/_/_/_/
\\
Trispectrum &
scalar anisotropic stress &
0.6 &
this work \\ [5 pt]
%_/_/_/_/_/_/_/_/_/_/_/_/_/_/_/_/_/_/_/_/_/_/_/_/_/_/_/_/_/_/_/_/
\\
Trispectrum &
magnetic inflationary mode &
0.05 &
this work; using \cite{bonvin13} \\ [5 pt]
%_/_/_/_/_/_/_/_/_/_/_/_/_/_/_/_/_/_/_/_/_/_/_/_/_/_/_/_/_/_/_/_/
\hline \hline
\end{tabular}
\label{B_0_limits}
\end{table}
% ==============================================================

\end{widetext}

\section{Conclusions}
\label{s:Concl}

We have presented the full calculation for the CMB trispectrum sourced by primordial magnetic field scalar modes, 
first reported in our Letter \cite{TSS11_Trispec_Letter}. In addition, we have calculated the scalar anisotropic stress
trispectrum in the flat-sky limit. Together with recent improved observational constraints on primordial non-Gaussianity from the \textit{Planck}
mission 2013 data, the magnetic scalar trispectrum enables us to place sub-nanoGauss upper limits on the strength primordial magnetic fields.

Magnetic energy density gives rise to a trispectrum of magnitude
$\approx 10^{-29}$, for $s$-independent terms. Also, the collinear configuration trispectrum for energy density, including all angular terms, 
gives a result that is very similar to the case of $s$-independent terms for energy density. 

For magnetic scalar anisotropic stress, we find a trispectrum of magnitude $ \approx 10^{-19}$, 
which is ten orders of magnitude larger than the magnetic energy density trispectrum. We also present an independent flat-sky limit
calculation of this trispectrum with its angular structure that yields a slightly larger trispectrum of magnitude $\approx 10^{-18}$.

The magnetic energy density trispectrum allows us to place stronger upper limits on the 
primordial magnetic field compared to a similar calculation with the magnetic energy density bispectrum \cite{SS09,Caprini09,Cai10}.
Further, the much larger trispectrum due to magnetic scalar anisotropic stress leads to the tightest constraint 
so far on large scale magnetic fields of $\sim$ 0.6 nG. 
This is approximately four times as strong as the corresponding upper limit from our previous bispectrum calculation ($\sim$ 2.4 nG) \cite{TSS10}. 
We note that the vector and tensor mode bispectra have been calculated numerically 
\cite{Shiraishi_vector,Shiraishi_tensor,Shiraishi_tensor_WMAP7} and give magnetic field limits of $\sim$ 3-10 nG. 
Recently, polarization bispectra \cite{Shiraishi_pol} constraints on magnetic fields have been forecast to be $\sim$ 2-3 nG from 
expected \textit{Planck} mission CMB polarization data. However, the scalar temperature trispectrum calculated in this work gives 
stronger magnetic fields constraints compared to the various kinds of bispectra that have been calculated (see Table(\ref{B_0_limits})).   
The trispectrum's sensitivity can be illustrated by the magnetic to inflationary scalar trispectrum ratio, which is $\sim 10^2$ 
compared to $\sim 0.1$ for the ratio of magnetic to inflationary scalar bispectra (taking $f_{NL}\sim10$ and $B_0 \sim 3 \text{ nG}$).

We also note that the magnetic field upper limit at megaparsec scales derived from just the scalar mode magnetic CMB 
trispectrum is already several times better than the upper limit from the magnetic CMB power spectrum combining 
scalar, vector and tensor modes: 3.4 nG from \textit{Planck} mission 2013 data \cite{Planck_Parameters} and 
($\sim$ 2-6 nG) from WMAP data \cite{Pao12,Pao10,Yam10,SL10}. Non-Gaussian correlations like the bispectrum and especially the trispectrum are
better able to constrain primordial cosmological magnetic fields than the CMB power spectrum. 

Finally, we have utilized the recently uncovered magnetic inflationary mode \cite{bonvin13} as a source for the CMB trispectrum. 
This new magnetic mode dominates over both energy density and scalar anisotropic stress and leads to an order of magnitude stronger constraint
on the primordial magnetic field of $\sim$ 0.05 nG. Further detailed investigation of the role this magnetic mode can play in sourcing 
various CMB correlations will be important.

Table (\ref{B_0_limits}) summarizes 
the current constraints on primordial magnetic fields derived from various probes using CMB anisotropies,
Thus, the CMB trispectrum is a new and more powerful probe of large scale primordial magnetic fields in the Universe.

Future consideration of magnetic vector and tensor modes in the trispectrum is likely to give additional constraints 
on primordial magnetic fields. Further improvement in magnetic field constraints 
is also possible from better $\tau_{NL}$ constraints that may emerge from a detailed analysis 
of the full \textit{Planck} mission data.

\section*{Acknowledgements}
PT and TRS would like to acknowledge the IUCAA Associateship Program as well as 
the facilities at the IUCAA Resource Center, University of Delhi. PT would like to acknowledge support 
from Sri Venkateswara College, University of Delhi, in pursuing this work. TRS acknowledges support 
from CSIR India via grant-in-aid no. 03(1187)/11/EMR-II.\\

\appendix*

\section{A}

In this Appendix we present the complete expressions for all angular terms generated by the sixteen operators present 
in the four-point correlation of magnetic anisotropic stress $\langle \Pi_B(\kk_1)\Pi_B(\kk_2)\Pi_B(\kk_3)\Pi_B(\kk_4)\rangle$ 
(Eq. \ref{16operators_Pi}).
The extensive angular term expressions presented below have also been checked by taking an alternative order of contraction 
while calculating angular terms.

Each operator term $X$ from 1 to 16 generates its own separate angular term expression ${\cal F}_{(I)}^{\fbox{\tiny{X}}}$. 
When summed over all $X$ this yields the angular term expression ${\cal F}_{(I)}$, 
where $I$ takes values 1 to 6 in the six term mode-coupling integral $\left[ \psi_{_{1234}}\right]_{\Pi}$ below. 

\begin{widetext}
\EQA
\psi_{_{1234}} = \frac{8}{(8\pi\rho_0)^4}
\int d^3 s M(s) M(\vert \kk_1 + \ssvec \vert) \Big[ 
  &M&(\left| \kk_1 + \kk_3 +\ssvec \right|) \left( M(\vert \kk_2  -  \ssvec \vert) {\cal F}_{(1)} 
                                                + M(\vert \kk_4  -  \ssvec \vert) {\cal F}_{(2)} \right) \nonumber \\
+ &M&(\left| \kk_1 + \kk_2 +\ssvec \right|) \left( M(\vert \kk_3  -  \ssvec \vert) {\cal F}_{(3)} 
                                                + M(\vert \kk_4  -  \ssvec \vert) {\cal F}_{(4)} \right) \nonumber \\
+ &M&(\left| \kk_1 + \kk_4 +\ssvec \right|) \left( M(\vert \kk_2  -  \ssvec \vert) {\cal F}_{(5)} 
                                                + M(\vert \kk_3  -  \ssvec \vert) {\cal F}_{(6)} \right) \Big]. \nonumber
% \label{psi}
\ENA
As seen in Eq. (\ref{16operators_Pi}), 
the angular term expressions $\mathcal{F}$ generated by operators $\fbox{2}$ to $\fbox{5}$ will carry a prefactor of (-3),
angular term expressions generated by $\fbox{6}$ to $\fbox{11}$ will carry a prefactor of (9), 
angular term expressions generated by $\fbox{12}$ to $\fbox{15}$ will carry a prefactor of (-27) and the 
angular term expressions generated by $\fbox{16}$ will have a prefactor of (81). 
For clarity, we suppress these prefactors while writing out the full angular term expressions below. 
The angles involved in these expressions have been defined earlier 
in Equations (\ref{phi_subscript_def}), (\ref{phi_def}) and in Table (\ref{Pi_angles_overbars}).

The angular terms for operator $\fbox{1}$ are
\EQA
 \mathcal{F}_{(1)}^{\fbox{\tiny{1}}}&=& -1 + \left( \alpha_1^2 + \alpha_2^2 +\alpha_6^2 + \beta_2^2 + \beta_6^2 + \gamma_6^2 \right)
     - \left( \alpha_1\alpha_2\beta_2 + \alpha_1\alpha_6\beta_6 + \alpha_2\alpha_6\gamma_6 + \beta_2\beta_6\gamma_6 \right)
     + \alpha_1\alpha_2\beta_6\gamma_6 \nonumber \\
 \mathcal{F}_{(2)}^{\fbox{\tiny{1}}}&=& -1 + \left( \alpha_1^2 + \alpha_4^2 +\alpha_6^2 + \beta_4^2 + \beta_6^2 + \epsilon_6^2 \right)
     - \left( \alpha_1\alpha_4\beta_4 + \alpha_1\alpha_6\beta_6 + \alpha_4\alpha_6\epsilon_6 + \beta_4\beta_6\epsilon_6 \right)
     + \alpha_1\alpha_4\beta_6\epsilon_6 \nonumber \\
 \mathcal{F}_{(3)}^{\fbox{\tiny{1}}}&=& -1 + \left( \alpha_1^2 + \alpha_3^2 +\alpha_5^2 + \beta_3^2 + \beta_5^2 + \delta_5^2 \right)
     - \left( \alpha_1\alpha_3\beta_3 + \alpha_1\alpha_5\beta_5 + \alpha_3\alpha_5\delta_5 + \beta_3\beta_5\delta_5 \right)
     + \alpha_1\alpha_3\beta_5\delta_5 \nonumber \\
 \mathcal{F}_{(4)}^{\fbox{\tiny{1}}}&=& -1 + \left( \alpha_1^2 + \alpha_4^2 +\alpha_5^2 + \beta_4^2 + \beta_5^2 + \epsilon_5^2 \right)
     - \left( \alpha_1\alpha_4\beta_4 + \alpha_1\alpha_5\beta_5 + \alpha_4\alpha_5\epsilon_5 + \beta_4\beta_5\epsilon_5 \right)
     + \alpha_1\alpha_4\beta_5\epsilon_5 \nonumber \\
 \mathcal{F}_{(5)}^{\fbox{\tiny{1}}}&=& -1 + \left( \alpha_1^2 + \alpha_2^2 +\alpha_7^2 + \beta_2^2 + \beta_7^2 + \gamma_7^2 \right)
     - \left( \alpha_1\alpha_2\beta_2 + \alpha_1\alpha_7\beta_7 + \alpha_2\alpha_7\gamma_7 + \beta_2\beta_7\gamma_7 \right)
     + \alpha_1\alpha_2\beta_7\gamma_7 \nonumber \\
 \mathcal{F}_{(6)}^{\fbox{\tiny{1}}}&=& -1 + \left( \alpha_1^2 + \alpha_3^2 +\alpha_7^2 + \beta_3^2 + \beta_7^2 + \delta_7^2 \right)
     - \left( \alpha_1\alpha_3\beta_3 + \alpha_1\alpha_7\beta_7 + \alpha_3\alpha_7\delta_7 + \beta_3\beta_7\delta_7 \right)
     + \alpha_1\alpha_3\beta_7\delta_7.
%  \label{72terms}
\ENA

%ooooooooooooooooooooooooooooooooooooooooooooooooooooooooooooooooooooooooooooooooooooooo

The angular terms for operator $\fbox{2}$ are
\EQA
 \mathcal{F}_{(1)}^{\fbox{\tiny{2}}} = &\,& \qquad \,\,\, 1 \quad \,\, - \overline{\alpha}_4^2 \quad -\overline{\beta}_4^2 \quad \!\!\!\! + \alpha_1\overline{\alpha}_4\overline{\beta}_4
 \quad - \quad \overline{\gamma}_4\left[ \overline{\gamma}_4 - \overline{\alpha}_4\alpha_2 - \overline{\beta}_4\beta_2 + \alpha_1\overline{\beta}_4\alpha_2 \right] \nonumber \\
&& - \; \overline{\lambda}_4\left[ \overline{\lambda}_4 - \overline{\alpha}_4\alpha_6 - \overline{\beta}_4\beta_6 + \alpha_1\overline{\alpha}_4\beta_6 \right] 
 \,+\,\,\, \overline{\gamma}_4\overline{\lambda}_4\left[ \gamma_6 - \alpha_2\alpha_6 - \beta_2\beta_6 + \alpha_1\alpha_2\beta_6 \right] \nonumber \\
%--------------------
 \mathcal{F}_{(2)}^{\fbox{\tiny{2}}} = &\,& \qquad \,\,\,  1 \quad \,\, - \overline{\beta}_4^2 \quad -\overline{\lambda}_4^2 \quad \!\!\!\! + \beta_6\overline{\lambda}_4\overline{\beta}_4
\quad - \quad \overline{\alpha}_4\left[ \overline{\alpha}_4 - \overline{\beta}_4\alpha_1 - \overline{\lambda}_4\alpha_6 + \beta_6\overline{\lambda}_4\alpha_1 \right] \nonumber \\
&& - \; \overline{\epsilon}_4\left[ \overline{\epsilon}_4 - \overline{\beta}_4\beta_4 - \overline{\lambda}_4\epsilon_6 + \beta_6\overline{\beta}_4\epsilon_6 \right]
\quad + \,\,\, \overline{\alpha}_4\overline{\epsilon}_4\left[ \alpha_4 - \alpha_1\beta_4 - \alpha_6\epsilon_6 + \beta_6\alpha_1\epsilon_6 \right] \nonumber \\
%--------------------
 \mathcal{F}_{(3)}^{\fbox{\tiny{2}}} = &\,& \qquad \,\,\, 1 \quad \,\, - \overline{\alpha}_4^2 \quad -\overline{\beta}_4^2 \quad \!\!\!\! + \alpha_1\overline{\alpha}_4\overline{\beta}_4
\quad - \quad \overline{\delta}_4\left[ \overline{\delta}_4 - \overline{\alpha}_4\alpha_3 - \overline{\beta}_4\beta_3 + \alpha_1\overline{\beta}_4\alpha_3 \right] \nonumber \\
&& - \; \overline{\kappa}_4\left[ \overline{\kappa}_4 - \overline{\alpha}_4\alpha_5 - \overline{\beta}_4\beta_5 + \alpha_1\overline{\alpha}_4\beta_5 \right] 
\, + \,\,\, \overline{\delta}_4\overline{\kappa}_4\left[ \delta_5 - \alpha_3\alpha_5 - \beta_3\beta_5 + \alpha_1\alpha_3\beta_5 \right] \nonumber \\
%--------------------
 \mathcal{F}_{(4)}^{\fbox{\tiny{2}}} = &\,& \qquad \,\,\, 1 \quad \,\, - \overline{\beta}_4^2 \quad -\overline{\kappa}_4^2 \quad \!\!\!\! + \beta_5\overline{\kappa}_4\overline{\beta}_4
\quad - \quad \overline{\alpha}_4\left[ \overline{\alpha}_4 - \overline{\beta}_4\alpha_1 - \overline{\kappa}_4\alpha_5+ \beta_5\overline{\kappa}_4\alpha_1 \right] \nonumber \\
&& - \; \overline{\epsilon}_4\left[ \overline{\epsilon}_4 - \overline{\beta_4}\beta_4 - \overline{\kappa}_4\epsilon_5 + \beta_5\overline{\beta}_4\epsilon_5 \right]
 \quad + \,\,\, \overline{\alpha}_4\overline{\epsilon}_4\left[ \alpha_4 - \alpha_1\beta_4 - \alpha_5\epsilon_5 + \beta_5\alpha_1\epsilon_5 \right] \nonumber \\
%--------------------
 \mathcal{F}_{(5)}^{\fbox{\tiny{2}}} = &\,& \qquad \,\,\, 1 \quad \,\, - \overline{\alpha}_4^2 \quad -\overline{\gamma}_4^2 \quad \!\!\!\! + \alpha_2\overline{\alpha}_4\overline{\gamma}_4
\quad - \quad \overline{\beta}_4\left[ \overline{\beta}_4 - \overline{\alpha}_4\alpha_1 - \overline{\gamma}_4\beta_2 + \alpha_2\overline{\gamma}_4\alpha_1 \right] \nonumber \\
&& - \; \overline{\chi}_4\left[ \overline{\chi}_4 - \overline{\alpha}_4\alpha_7 - \overline{\gamma}_4\gamma_7 + \alpha_2\overline{\alpha}_4\gamma_7 \right]
\,\,\, + \,\, \overline{\beta}_4\overline{\chi}_4\left[ \beta_7 - \alpha_1\alpha_7 - \beta_2\gamma_7 + \alpha_2\alpha_1\gamma_7 \right] \nonumber \\
%--------------------
 \mathcal{F}_{(6)}^{\fbox{\tiny{2}}} = &\,& \qquad \,\,\, 1 \quad \,\, - \overline{\alpha}_4^2 \quad -\overline{\delta}_4^2 \quad \!\!\!\! + \alpha_3\overline{\alpha}_4\overline{\delta}_4
\quad - \quad \overline{\beta}_4\left[ \overline{\beta}_4 - \overline{\alpha}_4\alpha_1 - \overline{\delta}_4\beta_3 + \alpha_3\overline{\delta}_4\alpha_1 \right] \nonumber \\
&& - \; \overline{\chi}_4\left[ \overline{\chi}_4 - \overline{\alpha}_4\alpha_7 - \overline{\delta}_4\delta_7 + \alpha_3\overline{\alpha}_4\delta_7 \right]
\, + \,\,\, \overline{\beta}_4\overline{\chi}_4\left[ \beta_7 - \alpha_1\alpha_7 - \beta_3\delta_7 + \alpha_3\alpha_1\delta_7 \right].
% \label{ang_terms_Pi_2}
\ENA

%ooooooooooooooooooooooooooooooooooooooooooooooooooooooooooooooooooooooooooooooooooooooo

The angular terms for operator $\fbox{3}$ are

\EQA
 \mathcal{F}_{(1)}^{\fbox{\tiny{3}}} = &\,& \qquad \,\,\, 1 \quad \,\, - \overline{\alpha}_3^2 \quad -\overline{\gamma}_3^2 \quad \!\!\!\! + \alpha_2\overline{\alpha}_3\overline{\gamma}_3
\quad - \quad \overline{\beta}_3\left[ \overline{\beta}_3 - \overline{\alpha}_3\alpha_1 - \overline{\gamma}_3\beta_2 + \alpha_2\overline{\gamma}_3\alpha_1 \right] \nonumber \\
&& - \; \overline{\lambda}_3\left[ \overline{\lambda}_3 - \overline{\alpha}_3\alpha_6 - \overline{\gamma}_3\gamma_6 + \alpha_2\overline{\alpha}_3\gamma_6 \right]
 \,\,\, + \,\, \overline{\beta}_3\overline{\lambda}_3\left[ \beta_6 - \alpha_1\alpha_6 - \beta_2\gamma_6 + \alpha_2\alpha_1\gamma_6 \right] \nonumber \\
%--------------------
  \mathcal{F}_{(2)}^{\fbox{\tiny{3}}} = &\,& \qquad \,\,\, 1 \quad \,\, - \overline{\alpha}_3^2 \quad -\overline{\epsilon}_3^2 \quad \!\!\!\! + \alpha_4\overline{\alpha}_3\overline{\epsilon}_3 
\quad - \quad \overline{\beta}_3\left[ \overline{\beta}_3 - \overline{\alpha}_3\alpha_1 - \overline{\epsilon}_3\beta_4 + \alpha_4\overline{\epsilon}_3\alpha_1 \right] \nonumber \\
&& - \; \overline{\lambda}_3\left[ \overline{\lambda}_3 - \overline{\alpha}_3\alpha_6 - \overline{\epsilon}_3\epsilon_6 + \alpha_4\overline{\alpha}_3\epsilon_6 \right] 
\,\,\, + \,\, \overline{\beta}_3\overline{\lambda}_3\left[ \beta_6 - \alpha_1\alpha_6 - \beta_4\epsilon_6 + \alpha_4\alpha_1\epsilon_6 \right] \nonumber \\
%--------------------
 \mathcal{F}_{(3)}^{\fbox{\tiny{3}}} = &\,& \qquad \,\,\, 1 \quad \,\, - \overline{\beta}_3^2 \quad -\overline{\kappa}_3^2 \quad \!\!\!\! + \beta_5\overline{\beta}_3\overline{\kappa}_3 
\quad \!- \quad \!\overline{\alpha}_3\left[ \overline{\alpha}_3 - \overline{\beta}_3\alpha_1 - \overline{\kappa}_3\alpha_5 + \beta_5\overline{\kappa}_3\alpha_1 \right] \nonumber \\
&& - \; \overline{\delta}_3\left[ \overline{\delta}_3 - \overline{\beta}_3\beta_3 - \overline{\kappa}_3\delta_5 + \beta_5\overline{\beta}_3\delta_5 \right] 
\,\,\,\,\, + \,\, \overline{\alpha}_3\overline{\delta}_3\left[ \alpha_3 - \alpha_1\beta_3 - \alpha_5\delta_5 + \beta_5\alpha_1\delta_5 \right] \nonumber \\
%--------------------
 \mathcal{F}_{(4)}^{\fbox{\tiny{3}}} = &\,& \qquad \,\,\, 1 \quad \,\, - \overline{\alpha}_3^2 \quad -\overline{\beta}_3^2 \quad \!\!\!\! + \alpha_1\overline{\alpha}_3\overline{\beta}_3
\quad - \quad \overline{\epsilon}_3\left[ \overline{\epsilon}_3 - \overline{\alpha}_3\alpha_4 - \overline{\beta}_3\beta_4 + \alpha_1\overline{\beta}_3\alpha_4 \right] \nonumber \\
&& - \; \overline{\kappa}_3\left[ \overline{\kappa}_3 - \overline{\alpha}_3\alpha_5 - \overline{\beta}_3\beta_5 + \alpha_1\overline{\alpha}_3\beta_5 \right]
\,\, + \,\, \overline{\epsilon}_3\overline{\kappa}_3\left[ \epsilon_5 - \alpha_4\alpha_5 - \beta_4\beta_5 + \alpha_1\alpha_4\beta_5 \right] \nonumber \\
%--------------------
 \mathcal{F}_{(5)}^{\fbox{\tiny{3}}} = &\,& \qquad \,\,\, 1 \quad \,\, - \overline{\alpha}_3^2 \quad -\overline{\beta}_3^2 \quad \!\!\!\! + \alpha_1\overline{\alpha}_3\overline{\beta}_3
\quad\! - \,\quad \overline{\gamma}_3\left[ \overline{\gamma}_3 - \overline{\alpha}_3\alpha_2 - \overline{\beta}_3\beta_2 + \alpha_1\overline{\beta}_3\alpha_2 \right] \nonumber \\
&& - \; \overline{\chi}_3\left[ \overline{\chi}_3 - \overline{\alpha}_3\alpha_7 - \overline{\beta}_3\beta_7 + \alpha_1\overline{\alpha}_3\beta_7 \right]
\,\, + \,\, \overline{\gamma}_3\overline{\chi}_3\left[ \gamma_7 - \alpha_2\alpha_7 - \beta_2\beta_7 + \alpha_1\alpha_2\beta_7 \right] \nonumber \\
%--------------------
 \mathcal{F}_{(6)}^{\fbox{\tiny{3}}} = &\,& \qquad \,\,\, 1 \quad \,\, - \overline{\beta}_3^2 \quad -\overline{\chi}_3^2 \quad \!\!\!\! + \beta_7\overline{\beta}_3\overline{\chi}_3
\quad \!\!- \quad \!\overline{\alpha}_3\left[ \overline{\alpha}_3 - \overline{\beta}_3\alpha_1 - \overline{\chi}_3\alpha_7 + \beta_7\overline{\chi}_3\alpha_1 \right] \nonumber \\
&& - \; \overline{\delta}_3\left[ \overline{\delta}_3 - \overline{\beta}_3\beta_3 - \overline{\chi}_3\delta_7 + \beta_7\overline{\beta}_3\delta_7 \right]
\,\,\, + \,\,\, \overline{\alpha}_3\overline{\delta}_3\left[ \alpha_3 - \alpha_1\beta_3 - \alpha_7\delta_7 + \beta_7\alpha_1\delta_7 \right].
% \label{ang_terms_Pi_2}
\ENA

%ooooooooooooooooooooooooooooooooooooooooooooooooooooooooooooooooooooooooooooooooooooooo

The angular terms for operator $\fbox{4}$ are
\EQA
 \mathcal{F}_{(1)}^{\fbox{\tiny{4}}} = &\,& \qquad \,\,\, 1 \quad \,\, - \overline{\beta}_2^2 \quad -\overline{\lambda}_2^2 \quad \!\!\!\! + \beta_6\overline{\beta}_2\overline{\lambda}_2
\quad \!- \,\quad \overline{\alpha}_2\left[ \overline{\alpha}_2 - \overline{\beta}_2\alpha_1 - \overline{\lambda}_2\alpha_6 + \beta_6\overline{\lambda}_2\alpha_1 \right] \nonumber \\
&& - \; \overline{\gamma}_2\left[ \overline{\gamma}_2 - \overline{\beta}_2\beta_2 - \overline{\lambda}_2\gamma_6 + \beta_6\overline{\beta}_2\gamma_6 \right]
\,\,\, + \,\,\, \overline{\alpha}_2\overline{\gamma}_2\left[ \alpha_2 - \alpha_1\beta_2 - \alpha_6\gamma_6 + \beta_6\alpha_1\gamma_6 \right] \nonumber \\
%--------------------
 \mathcal{F}_{(2)}^{\fbox{\tiny{4}}} = &\,& \qquad \,\,\, 1 \quad \,\, - \overline{\alpha}_2^2 \quad -\overline{\beta}_2^2 \quad \!\!\!\! + \alpha_1\overline{\alpha}_2\overline{\beta}_2
\quad \!- \,\quad \overline{\epsilon}_2\left[ \overline{\epsilon}_2 - \overline{\alpha}_2\alpha_4 - \overline{\beta}_2\beta_4 + \alpha_1\overline{\beta}_2\alpha_4 \right] \nonumber \\
&& - \; \overline{\lambda}_2\left[ \overline{\lambda}_2 - \overline{\alpha}_2\alpha_6 - \overline{\beta}_2\beta_6 + \alpha_1\overline{\alpha}_2\beta_6 \right]
\,\, + \,\, \overline{\epsilon}_2\overline{\lambda}_2\left[ \epsilon_6 - \alpha_4\alpha_6 - \beta_4\beta_6 + \alpha_1\alpha_4\beta_6 \right] \nonumber \\
%--------------------
 \mathcal{F}_{(3)}^{\fbox{\tiny{4}}} = &\,& \qquad \,\,\, 1 \quad \,\, - \overline{\alpha}_2^2 \quad -\overline{\delta}_2^2 \quad \!\!\!\! + \alpha_3\overline{\alpha}_2\overline{\delta}_2
\quad \!\!- \,\,\quad \overline{\beta}_2\left[ \overline{\beta}_2 - \overline{\alpha}_2\alpha_1 - \overline{\delta}_2\beta_3 + \alpha_3\overline{\delta}_2\alpha_1 \right] \nonumber \\
&& - \; \overline{\kappa}_2\left[ \overline{\kappa}_2 - \overline{\alpha}_2\alpha_5 - \overline{\delta}_2\delta_5 + \alpha_3\overline{\alpha}_2\delta_5 \right]
\,\, + \,\, \overline{\beta}_2\overline{\kappa}_2\left[ \beta_5 - \alpha_1\alpha_5 - \beta_3\delta_5 + \alpha_3\alpha_1\delta_5 \right] \nonumber \\
%--------------------
 \mathcal{F}_{(4)}^{\fbox{\tiny{4}}} = &\,& \qquad \,\,\, 1 \quad \,\, - \overline{\alpha}_2^2 \quad -\overline{\epsilon}_2^2 \quad \!\!\!\! + \alpha_4\overline{\alpha}_2\overline{\epsilon}_2
\quad \!\!- \,\quad \overline{\beta}_2\left[ \overline{\beta}_2 - \overline{\alpha}_2\alpha_1 - \overline{\epsilon}_2\beta_4 + \alpha_4\overline{\epsilon}_2\alpha_1 \right] \nonumber \\
&& - \; \overline{\kappa}_2\left[ \overline{\kappa}_2 - \overline{\alpha}_2\alpha_5 - \overline{\epsilon}_2\epsilon_5 + \alpha_4\overline{\alpha}_2\epsilon_5 \right]
\,\, + \,\, \overline{\beta}_2\overline{\kappa}_2\left[ \beta_5 - \alpha_1\alpha_5 - \beta_4\epsilon_5 + \alpha_4\alpha_1\epsilon_5 \right] \nonumber \\
%--------------------
 \mathcal{F}_{(5)}^{\fbox{\tiny{4}}} = &\,& \qquad \,\,\, 1 \quad \,\, - \overline{\beta}_2^2 \quad -\overline{\chi}_2^2 \quad \!\!\!\! + \beta_7\overline{\beta}_2\overline{\chi}_2
\quad \!\!- \,\,\quad \overline{\alpha}_2\left[ \overline{\alpha}_2 - \overline{\beta}_2\alpha_1 - \overline{\chi}_2\alpha_7 + \beta_7\overline{\chi}_2\alpha_1 \right] \nonumber \\
&& - \; \overline{\gamma}_2\left[ \overline{\gamma}_2 - \overline{\beta}_2\beta_2 - \overline{\chi}_2\gamma_7 + \beta_7\overline{\beta}_2\gamma_7 \right]
\,\, + \,\, \overline{\alpha}_2\overline{\gamma}_2\left[ \alpha_2 - \alpha_1\beta_2 - \alpha_7\gamma_7 + \beta_7\alpha_1\gamma_7 \right] \nonumber \\
%--------------------
 \mathcal{F}_{(6)}^{\fbox{\tiny{4}}} = &\,& \qquad \,\,\, 1 \quad \,\, - \overline{\alpha}_2^2 \quad -\overline{\beta}_2^2 \quad \!\!\!\! + \alpha_1\overline{\alpha}_2\overline{\beta}_2
\quad - \quad \overline{\delta}_2\left[ \overline{\delta}_2 - \overline{\alpha}_2\alpha_3 - \overline{\beta}_2\beta_3 + \alpha_1\overline{\beta}_2\alpha_3 \right] \nonumber \\
&& - \; \overline{\chi}_2\left[ \overline{\chi}_2 - \overline{\alpha}_2\alpha_7 - \overline{\beta}_2\beta_7 + \alpha_1\overline{\alpha}_2\beta_7 \right]
\,\, + \,\, \overline{\delta}_2\overline{\chi}_2\left[ \delta_7 - \alpha_3\alpha_7 - \beta_3\beta_7 + \alpha_1\alpha_3\beta_7 \right].
% \label{ang_terms_Pi_2}
\ENA

%ooooooooooooooooooooooooooooooooooooooooooooooooooooooooooooooooooooooooooooooooooooooo

The angular terms for operator $\fbox{5}$ are
\EQA
 \mathcal{F}_{(1)}^{\fbox{\tiny{5}}} = &\,& \qquad \,\,\, 1 \quad \,\, - \overline{\gamma}_1^2 \quad -\overline{\lambda}_1^2 \quad \!\!\!\! + \gamma_6\overline{\gamma}_1\overline{\lambda}_1
\quad - \quad \overline{\alpha}_1\left[ \overline{\alpha}_1 - \overline{\gamma}_1\alpha_2 - \overline{\lambda}_1\alpha_6 + \gamma_6\overline{\lambda}_1\alpha_2 \right] \nonumber \\
&& - \; \overline{\beta}_1\left[ \overline{\beta}_1 - \overline{\gamma}_1\beta_2 - \overline{\lambda}_1\beta_6 + \gamma_6\overline{\gamma}_1\beta_6 \right]
\,\, + \,\, \overline{\alpha}_1\overline{\beta}_1\left[ \alpha_1 - \alpha_2\beta_2 - \alpha_6\beta_6 + \gamma_6\alpha_2\beta_6 \right] \nonumber \\
%--------------------
 \mathcal{F}_{(2)}^{\fbox{\tiny{5}}} = &\,& \qquad \,\,\, 1 \quad \,\, - \overline{\epsilon}_1^2 \quad -\overline{\lambda}_1^2 \quad \!\!\!\! + \epsilon_6\overline{\epsilon}_1\overline{\lambda}_1
\quad - \quad \overline{\alpha}_1\left[ \overline{\alpha}_1 - \overline{\epsilon}_1\alpha_4 - \overline{\lambda}_1\alpha_6 + \epsilon_6\overline{\lambda}_1\alpha_4 \right] \nonumber \\
&& - \; \overline{\beta}_1\left[ \overline{\beta}_1 - \overline{\epsilon}_1\beta_4 - \overline{\lambda}_1\beta_6 + \epsilon_6\overline{\epsilon}_1\beta_6 \right] 
\,\, + \,\, \overline{\alpha}_1\overline{\beta}_1\left[ \alpha_1 - \alpha_4\beta_4 - \alpha_6\beta_6 + \epsilon_6\alpha_4\beta_6 \right] \nonumber \\
%--------------------
 \mathcal{F}_{(3)}^{\fbox{\tiny{5}}} = &\,& \qquad \,\,\, 1 \quad \,\, - \overline{\delta}_1^2 \quad -\overline{\kappa}_1^2 \quad \!\!\!\! + \delta_5\overline{\delta}_1\overline{\kappa}_1 
\quad - \quad \overline{\alpha}_1\left[ \overline{\alpha}_1 - \overline{\delta}_1\alpha_3 - \overline{\kappa}_1\alpha_5 + \delta_5\overline{\kappa}_1\alpha_3 \right] \nonumber \\
&& - \; \overline{\beta}_1\left[ \overline{\beta}_1 - \overline{\delta}_1\beta_3 - \overline{\kappa}_1\beta_5 + \delta_5\overline{\delta}_1\beta_5 \right] 
\,\, + \,\, \overline{\alpha}_1\overline{\beta}_1\left[ \alpha_1 - \alpha_3\beta_3 - \alpha_5\beta_5 + \delta_5\alpha_3\beta_5 \right] \nonumber \\
%--------------------
 \mathcal{F}_{(4)}^{\fbox{\tiny{5}}} = &\,& \qquad \,\,\, 1 \quad \,\, - \overline{\epsilon}_1^2 \quad -\overline{\kappa}_1^2 \quad \!\!\!\! + \epsilon_5\overline{\epsilon}_1\overline{\kappa}_1 
\quad - \quad \,\overline{\alpha}_1\left[ \overline{\alpha}_1 - \overline{\epsilon}_1\alpha_4 - \overline{\kappa}_1\alpha_5 + \epsilon_5\overline{\kappa}_1\alpha_4 \right] \nonumber \\
&& - \; \overline{\beta}_1\left[ \overline{\beta}_1 - \overline{\epsilon}_1\beta_4 - \overline{\kappa}_1\beta_5 + \epsilon_5\overline{\epsilon}_1\beta_5 \right] 
\,\, + \,\, \overline{\alpha}_1\overline{\beta}_1\left[ \alpha_1 - \alpha_4\beta_4 - \alpha_5\beta_5 + \epsilon_5\alpha_4\beta_5 \right] \nonumber \\
%--------------------
 \mathcal{F}_{(5)}^{\fbox{\tiny{5}}} = &\,& \qquad \,\,\, 1 \quad \,\, - \overline{\gamma}_1^2 \quad -\overline{\chi}_1^2 \quad \!\!\!\! + \gamma_7\overline{\gamma}_1\overline{\chi}_1 
\quad \!- \,\quad \,\overline{\alpha}_1\left[ \overline{\alpha}_1 - \overline{\gamma}_1\alpha_2 - \overline{\chi}_1\alpha_7 + \gamma_7\overline{\chi}_1\alpha_2 \right] \nonumber \\
&& - \; \overline{\beta}_1\left[ \overline{\beta}_1 - \overline{\gamma}_1\beta_2 - \overline{\chi}_1\beta_7 + \gamma_7\overline{\gamma}_1\beta_7 \right] 
\,\, + \,\, \overline{\alpha}_1\overline{\beta}_1\left[ \alpha_1 - \alpha_2\beta_2 - \alpha_7\beta_7 + \gamma_7\alpha_2\beta_7 \right] \nonumber \\
%--------------------
 \mathcal{F}_{(6)}^{\fbox{\tiny{5}}} = &\,& \qquad \,\,\, 1 \quad \,\, - \overline{\delta}_1^2 \quad -\overline{\chi}_1^2 \quad \!\!\!\! + \delta_7\overline{\delta}_1\overline{\chi}_1 
\quad - \quad \overline{\alpha}_1\left[ \overline{\alpha}_1 - \overline{\delta}_1\alpha_3 - \overline{\chi}_1\alpha_7 + \delta_7\overline{\chi}_1\alpha_3 \right] \nonumber \\
&& - \; \overline{\beta}_1\left[ \overline{\beta}_1 - \overline{\delta}_1\beta_3 - \overline{\chi}_1\beta_7 + \delta_7\overline{\delta}_1\beta_7 \right] 
\,\, + \,\, \overline{\alpha}_1\overline{\beta}_1\left[ \alpha_1 - \alpha_3\beta_3 - \alpha_7\beta_7 + \delta_7\alpha_3\beta_7 \right]. 
% \label{ang_terms_Pi_2}
\ENA

%ooooooooooooooooooooooooooooooooooooooooooooooooooooooooooooooooooooooooooooooooooooooo

The angular terms for operator $\fbox{6}$ are

\EQA
\mathcal{F}_{(1)}^{\fbox{\tiny{6}}} &=&  
 \left[ \theta_{34} - \overline{\beta}_3\overline{\beta}_4  - \overline{\gamma}_4\left( \overline{\gamma}_3 -\beta_2\overline{\beta}_3 \right)
            - \overline{\alpha}_4\left( \overline{\alpha}_3 -\alpha_1\overline{\beta}_3 \right) 
    + \overline{\gamma}_4\alpha_2\left( \overline{\alpha}_3 -\alpha_1\overline{\beta}_3 \right) \right]
 \left( \theta_{34} - \overline{\lambda}_3\overline{\lambda}_4  \right)  \nonumber \\
%--------------------
\mathcal{F}_{(2)}^{\fbox{\tiny{6}}} &=&  
 \left( \theta_{34} - \overline{\alpha}_3\overline{\alpha}_4  - \overline{\beta}_3\overline{\beta}_4 + \alpha_1\overline{\alpha}_4 \overline{\beta}_3 \right)
 \left( \theta_{34} - \overline{\epsilon}_3\overline{\epsilon}_4  - \overline{\lambda}_3\overline{\lambda}_4 + \epsilon_6\overline{\epsilon}_4 \overline{\lambda}_3 \right) \nonumber \\
%--------------------
\mathcal{F}_{(3)}^{\fbox{\tiny{6}}} &=&  
 \left[ \theta_{34} - \overline{\alpha}_3\overline{\alpha}_4  - \overline{\kappa}_4\left( \overline{\kappa}_3 -\alpha_5\overline{\alpha}_3 \right)
            - \overline{\beta}_4\left( \overline{\beta}_3 -\alpha_1\overline{\alpha}_3 \right) 
    + \overline{\kappa}_4\beta_5\left( \overline{\beta}_3 -\alpha_1\overline{\alpha}_3 \right) \right]
 \left( \theta_{34} - \overline{\delta}_3\overline{\delta}_4  \right)  \nonumber \\
%--------------------
\mathcal{F}_{(4)}^{\fbox{\tiny{6}}} &=&  
 \left[ \theta_{34} - \overline{\alpha}_3\overline{\alpha}_4  - \overline{\kappa}_3\left( \overline{\kappa}_4 -\alpha_5\overline{\alpha}_4 \right)
            - \overline{\beta}_3\left( \overline{\beta}_4 -\alpha_1\overline{\alpha}_4 \right) 
    + \overline{\kappa}_3\beta_5\left( \overline{\beta}_4 -\alpha_1\overline{\alpha}_4 \right) \right]
 \left( \theta_{34} - \overline{\epsilon}_3\overline{\epsilon}_4  \right)  \nonumber \\
%--------------------
\mathcal{F}_{(5)}^{\fbox{\tiny{6}}} &=&  
 \left[ \theta_{34} - \overline{\beta}_3\overline{\beta}_4  - \overline{\gamma}_3\left( \overline{\gamma}_4 -\beta_2\overline{\beta}_4 \right)
            - \overline{\alpha}_3\left( \overline{\alpha}_4 -\alpha_1\overline{\beta}_4 \right) 
    + \overline{\gamma}_3\alpha_2\left( \overline{\alpha}_4 -\alpha_1\overline{\beta}_4 \right) \right]
 \left( \theta_{34} - \overline{\chi}_3\overline{\chi}_4  \right)  \nonumber \\
%--------------------
\mathcal{F}_{(6)}^{\fbox{\tiny{6}}} &=&  
 \left( \theta_{34} - \overline{\alpha}_3\overline{\alpha}_4  - \overline{\beta}_3\overline{\beta}_4 + \alpha_1\overline{\alpha}_3 \overline{\beta}_4 \right)
 \left( \theta_{34} - \overline{\delta}_3\overline{\delta}_4  - \overline{\chi}_3\overline{\chi}_4 + \delta_7\overline{\delta}_3 \overline{\chi}_4 \right).
\ENA

%ooooooooooooooooooooooooooooooooooooooooooooooooooooooooooooooooooooooooooooooooooooooo

The angular terms for operator $\fbox{7}$ are

\EQA
\mathcal{F}_{(1)}^{\fbox{\tiny{7}}} &=&  
 \left[ \theta_{24} - \overline{\alpha}_2\overline{\alpha}_4  - \overline{\lambda}_4\left( \overline{\lambda}_2 -\alpha_6\overline{\alpha}_2 \right)
            - \overline{\beta}_4\left( \overline{\beta}_2 -\alpha_1\overline{\alpha}_2 \right) 
    + \overline{\lambda}_4\beta_6\left( \overline{\beta}_2 -\alpha_1\overline{\alpha}_2 \right) \right]
 \left( \theta_{24} - \overline{\gamma}_2\overline{\gamma}_4  \right)  \nonumber \\
%--------------------
\mathcal{F}_{(2)}^{\fbox{\tiny{7}}} &=&  
 \left[ \theta_{24} - \overline{\alpha}_2\overline{\alpha}_4  - \overline{\lambda}_2\left( \overline{\lambda}_4 -\alpha_6\overline{\alpha}_4 \right)
            - \overline{\beta}_2\left( \overline{\beta}_4 -\alpha_1\overline{\alpha}_4 \right) 
    + \overline{\lambda}_2\beta_6\left( \overline{\beta}_4 -\alpha_1\overline{\alpha}_4 \right) \right]
 \left( \theta_{24} - \overline{\epsilon}_2\overline{\epsilon}_4  \right)  \nonumber \\
%--------------------
\mathcal{F}_{(3)}^{\fbox{\tiny{7}}} &=&  
 \left[ \theta_{24} - \overline{\beta}_2\overline{\beta}_4  - \overline{\delta}_4\left( \overline{\delta}_2 -\beta_3\overline{\beta}_2 \right)
            - \overline{\alpha}_4\left( \overline{\alpha}_2 -\alpha_1\overline{\beta}_2 \right) 
    + \overline{\delta}_4\alpha_3\left( \overline{\alpha}_2 -\alpha_1\overline{\beta}_2 \right) \right]
 \left( \theta_{24} - \overline{\kappa}_2\overline{\kappa}_4  \right)  \nonumber \\
%--------------------
\mathcal{F}_{(4)}^{\fbox{\tiny{7}}} &=&  
 \left( \theta_{24} - \overline{\alpha}_2\overline{\alpha}_4  - \overline{\beta}_2\overline{\beta}_4 + \alpha_1\overline{\alpha}_4 \overline{\beta}_2 \right)
 \left( \theta_{24} - \overline{\epsilon}_2\overline{\epsilon}_4  - \overline{\kappa}_2\overline{\kappa}_4 + \epsilon_5\overline{\epsilon}_4 \overline{\kappa}_2 \right) \nonumber \\
%--------------------
\mathcal{F}_{(5)}^{\fbox{\tiny{7}}} &=&  
 \left( \theta_{24} - \overline{\alpha}_2\overline{\alpha}_4  - \overline{\beta}_2\overline{\beta}_4 + \alpha_1\overline{\alpha}_2 \overline{\beta}_4 \right)
 \left( \theta_{24} - \overline{\gamma}_2\overline{\gamma}_4  - \overline{\chi}_2\overline{\chi}_4 + \gamma_7\overline{\gamma}_2 \overline{\chi}_4 \right) \nonumber \\
%--------------------
\mathcal{F}_{(6)}^{\fbox{\tiny{7}}} &=&  
 \left[ \theta_{24} - \overline{\beta}_2\overline{\beta}_4  - \overline{\delta}_2\left( \overline{\delta}_4 -\beta_3\overline{\beta}_4 \right)
            - \overline{\alpha}_2\left( \overline{\alpha}_4 -\alpha_1\overline{\beta}_4 \right) 
    + \overline{\delta}_2\alpha_3\left( \overline{\alpha}_4 -\alpha_1\overline{\beta}_4 \right) \right]
 \left( \theta_{24} - \overline{\chi}_2\overline{\chi}_4  \right).
\ENA

%ooooooooooooooooooooooooooooooooooooooooooooooooooooooooooooooooooooooooooooooooooooooo

The angular terms for operator $\fbox{8}$ are

\EQA
\mathcal{F}_{(1)}^{\fbox{\tiny{8}}} &=&  
 \left( \theta_{23} - \overline{\alpha}_2\overline{\alpha}_3  - \overline{\beta}_2\overline{\beta}_3 + \alpha_1\overline{\alpha}_2 \overline{\beta}_3 \right)
 \left( \theta_{23} - \overline{\gamma}_2\overline{\gamma}_3  - \overline{\lambda}_2\overline{\lambda}_3 + \gamma_6\overline{\gamma}_2 \overline{\lambda}_3 \right) \nonumber \\
%--------------------
\mathcal{F}_{(2)}^{\fbox{\tiny{8}}} &=&  
 \left[ \theta_{23} - \overline{\beta}_2\overline{\beta}_3  - \overline{\epsilon}_2\left( \overline{\epsilon}_3 -\beta_4\overline{\beta}_3 \right)
            - \overline{\alpha}_2\left( \overline{\alpha}_3 -\alpha_1\overline{\beta}_3 \right) 
    + \overline{\epsilon}_2\alpha_4\left( \overline{\alpha}_3 -\alpha_1\overline{\beta}_3 \right) \right]
 \left( \theta_{23} - \overline{\lambda}_2\overline{\lambda}_3  \right) \nonumber \\
%--------------------
\mathcal{F}_{(3)}^{\fbox{\tiny{8}}} &=&  
 \left( \theta_{23} - \overline{\alpha}_2\overline{\alpha}_3  - \overline{\beta}_2\overline{\beta}_3 + \alpha_1\overline{\alpha}_3 \overline{\beta}_2 \right)
 \left( \theta_{23} - \overline{\delta}_2\overline{\delta}_3  - \overline{\kappa}_2\overline{\kappa}_3 + \delta_5\overline{\delta}_3 \overline{\kappa}_2 \right) \nonumber \\
%--------------------
\mathcal{F}_{(4)}^{\fbox{\tiny{8}}} &=&  
 \left[ \theta_{23} - \overline{\beta}_2\overline{\beta}_3  - \overline{\epsilon}_3\left( \overline{\epsilon}_2 -\beta_4\overline{\beta}_2 \right)
            - \overline{\alpha}_3\left( \overline{\alpha}_2 -\alpha_1\overline{\beta}_2 \right) 
    + \overline{\epsilon}_3\alpha_4\left( \overline{\alpha}_2 -\alpha_1\overline{\beta}_2 \right) \right]
 \left( \theta_{23} - \overline{\kappa}_2\overline{\kappa}_3  \right) \nonumber \\
%--------------------
\mathcal{F}_{(5)}^{\fbox{\tiny{8}}} &=&  
 \left[ \theta_{23} - \overline{\alpha}_2\overline{\alpha}_3  - \overline{\chi}_3\left( \overline{\chi}_2 -\alpha_7\overline{\alpha}_2 \right)
            - \overline{\beta}_3\left( \overline{\beta}_2 -\alpha_1\overline{\alpha}_2 \right) 
    + \overline{\chi}_3\beta_7\left( \overline{\beta}_2 -\alpha_1\overline{\alpha}_2 \right) \right]
 \left( \theta_{23} - \overline{\gamma}_2\overline{\gamma}_3  \right)  \nonumber \\
%--------------------
\mathcal{F}_{(6)}^{\fbox{\tiny{8}}} &=&  
 \left[ \theta_{23} - \overline{\alpha}_2\overline{\alpha}_3  - \overline{\chi}_2\left( \overline{\chi}_3 -\alpha_7\overline{\alpha}_3 \right)
            - \overline{\beta}_2\left( \overline{\beta}_3 -\alpha_1\overline{\alpha}_3 \right) 
    + \overline{\chi}_2\beta_7\left( \overline{\beta}_3 -\alpha_1\overline{\alpha}_3 \right) \right]
 \left( \theta_{23} - \overline{\delta}_2\overline{\delta}_3  \right).
\ENA

%ooooooooooooooooooooooooooooooooooooooooooooooooooooooooooooooooooooooooooooooooooooooo

The angular terms for operator $\fbox{9}$ are

\EQA
\mathcal{F}_{(1)}^{\fbox{\tiny{9}}} &=&  
 \left( \theta_{14} - \overline{\alpha}_1\overline{\alpha}_4  - \overline{\gamma}_1\overline{\gamma}_4 + \alpha_2\overline{\alpha}_1 \overline{\gamma}_4 \right)
 \left( \theta_{14} - \overline{\beta}_1\overline{\beta}_4 - \overline{\lambda}_1\overline{\lambda}_4 + \alpha_6\overline{\beta}_1 \overline{\lambda}_4 \right) \nonumber \\
%--------------------
\mathcal{F}_{(2)}^{\fbox{\tiny{9}}} &=&  
 \left[ \theta_{14} - \overline{\beta}_1\overline{\beta}_4  - \overline{\epsilon}_4\left( \overline{\epsilon}_1 -\beta_4\overline{\beta}_1 \right)
            - \overline{\lambda}_4\left( \overline{\lambda}_1 -\beta_6\overline{\beta}_1 \right) 
    + \overline{\epsilon}_4\epsilon_6\left( \overline{\lambda}_1 -\beta_6\overline{\beta}_1 \right) \right]
 \left( \theta_{14} - \overline{\alpha}_1\overline{\alpha}_4  \right) \nonumber \\
%--------------------
\mathcal{F}_{(3)}^{\fbox{\tiny{9}}} &=&  
 \left( \theta_{14} - \overline{\alpha}_1\overline{\alpha}_4  - \overline{\delta}_1\overline{\delta}_4 + \alpha_2\overline{\alpha}_1 \overline{\delta}_4 \right)
 \left( \theta_{14} - \overline{\beta}_1\overline{\beta}_4  - \overline{\kappa}_1\overline{\kappa}_4 + \beta_5\overline{\beta}_1 \overline{\kappa}_4 \right) \nonumber \\
%--------------------
\mathcal{F}_{(4)}^{\fbox{\tiny{9}}} &=&  
 \left[ \theta_{14} - \overline{\beta}_1\overline{\beta}_4  - \overline{\epsilon}_4\left( \overline{\epsilon}_1 -\beta_4\overline{\beta}_1 \right)
            - \overline{\kappa}_4\left( \overline{\kappa}_1 -\beta_5\overline{\beta}_1 \right) 
    + \overline{\epsilon}_4\epsilon_5\left( \overline{\kappa}_1 -\beta_5\overline{\beta}_1 \right) \right]
 \left( \theta_{14} - \overline{\alpha}_1\overline{\alpha}_4  \right) \nonumber \\
%--------------------
\mathcal{F}_{(5)}^{\fbox{\tiny{9}}} &=&  
 \left[ \theta_{14} - \overline{\alpha}_1\overline{\alpha}_4  - \overline{\chi}_4\left( \overline{\chi}_1 -\alpha_7\overline{\alpha}_1 \right)
            - \overline{\gamma}_4\left( \overline{\gamma}_1 -\alpha_2\overline{\alpha}_1 \right) 
    + \overline{\chi}_4\gamma_7\left( \overline{\gamma}_1 -\alpha_2\overline{\alpha}_1 \right) \right]
 \left( \theta_{14} - \overline{\beta}_1\overline{\beta}_4  \right) \nonumber \\
%--------------------
\mathcal{F}_{(6)}^{\fbox{\tiny{9}}} &=&  
 \left[ \theta_{14} - \overline{\alpha}_1\overline{\alpha}_4  - \overline{\chi}_4\left( \overline{\chi}_1 -\alpha_7\overline{\alpha}_1 \right)
            - \overline{\delta}_4\left( \overline{\delta}_1 -\alpha_3\overline{\alpha}_1 \right) 
    + \overline{\chi}_4\delta_7\left( \overline{\delta}_1 -\alpha_3\overline{\alpha}_1 \right) \right]
 \left( \theta_{14} - \overline{\beta}_1\overline{\beta}_4  \right).
\ENA
%ooooooooooooooooooooooooooooooooooooooooooooooooooooooooooooooooooooooooooooooooooooooo

The angular terms for operator $\fbox{10}$ are

\EQA
\mathcal{F}_{(1)}^{\fbox{\tiny{10}}} &=&  
 \left[ \theta_{13} - \overline{\alpha}_1\overline{\alpha}_3  - \overline{\lambda}_3\left( \overline{\lambda}_1 -\alpha_6\overline{\alpha}_1 \right)
            - \overline{\gamma}_3\left( \overline{\gamma}_1 -\alpha_2\overline{\alpha}_1 \right) 
    + \overline{\lambda}_3\gamma_6\left( \overline{\gamma}_1 -\alpha_2\overline{\alpha}_1 \right) \right]
 \left( \theta_{13} - \overline{\beta}_1\overline{\beta}_3  \right) \nonumber \\
%--------------------
\mathcal{F}_{(2)}^{\fbox{\tiny{10}}} &=&  
 \left[ \theta_{13} - \overline{\alpha}_1\overline{\alpha}_3  - \overline{\lambda}_3\left( \overline{\lambda}_1 -\alpha_6\overline{\alpha}_1 \right)
            - \overline{\epsilon}_3\left( \overline{\epsilon}_1 -\alpha_4\overline{\alpha}_1 \right) 
    + \overline{\lambda}_3\epsilon_6\left( \overline{\epsilon}_1 -\alpha_4\overline{\alpha}_1 \right) \right]
 \left( \theta_{13} - \overline{\beta}_1\overline{\beta}_3  \right) \nonumber \\
%--------------------
\mathcal{F}_{(3)}^{\fbox{\tiny{10}}} &=&  
 \left[ \theta_{13} - \overline{\beta}_1\overline{\beta}_3  - \overline{\delta}_3\left( \overline{\delta}_1 -\beta_3\overline{\beta}_1 \right)
            - \overline{\kappa}_3\left( \overline{\kappa}_1 -\beta_5\overline{\beta}_1 \right) 
    + \overline{\delta}_3\delta_5\left( \overline{\kappa}_1 -\beta_5\overline{\beta}_1 \right) \right]
 \left( \theta_{13} - \overline{\alpha}_1\overline{\alpha}_3  \right) \nonumber \\
%--------------------
\mathcal{F}_{(4)}^{\fbox{\tiny{10}}} &=&  
 \left( \theta_{13} - \overline{\alpha}_1\overline{\alpha}_3  - \overline{\epsilon}_1\overline{\epsilon}_3 + \alpha_4\overline{\alpha}_1 \overline{\epsilon}_3 \right)
 \left( \theta_{13} - \overline{\beta}_1\overline{\beta}_3  - \overline{\kappa}_1\overline{\kappa}_3 + \beta_5\overline{\beta}_1 \overline{\kappa}_3 \right) \nonumber \\
%--------------------
\mathcal{F}_{(5)}^{\fbox{\tiny{10}}} &=&  
 \left( \theta_{13} - \overline{\alpha}_1\overline{\alpha}_3  - \overline{\gamma}_1\overline{\gamma}_3 + \alpha_2\overline{\alpha}_1 \overline{\gamma}_3 \right)
 \left( \theta_{13} - \overline{\beta}_1\overline{\beta}_3  - \overline{\chi}_1\overline{\chi}_3 + \beta_7\overline{\beta}_1 \overline{\chi}_3 \right) \nonumber \\
%--------------------
\mathcal{F}_{(6)}^{\fbox{\tiny{10}}} &=&  
 \left[ \theta_{13} - \overline{\beta}_1\overline{\beta}_3  - \overline{\delta}_3\left( \overline{\delta}_1 -\beta_3\overline{\beta}_1 \right)
            - \overline{\chi}_3\left( \overline{\chi}_1 -\beta_7\overline{\beta}_1 \right) 
    + \overline{\delta}_3\delta_7\left( \overline{\chi}_1 -\beta_7\overline{\beta}_1 \right) \right]
 \left( \theta_{13} - \overline{\alpha}_1\overline{\alpha}_3  \right). 
\ENA

%ooooooooooooooooooooooooooooooooooooooooooooooooooooooooooooooooooooooooooooooooooooooo

The angular terms for operator $\fbox{11}$ are

\EQA
\mathcal{F}_{(1)}^{\fbox{\tiny{11}}} &=&  
 \left[ \theta_{12} - \overline{\beta}_1\overline{\beta}_2  - \overline{\gamma}_2\left( \overline{\gamma}_1 -\beta_2\overline{\beta}_1 \right)
            - \overline{\lambda}_2\left( \overline{\lambda}_1 -\beta_6\overline{\beta}_1 \right) 
    + \overline{\gamma}_2\gamma_6\left( \overline{\lambda}_1 -\beta_6\overline{\beta}_1 \right) \right]
 \left( \theta_{12} - \overline{\alpha}_1\overline{\alpha}_2  \right) \nonumber \\
%--------------------
\mathcal{F}_{(2)}^{\fbox{\tiny{11}}} &=&  
 \left( \theta_{12} - \overline{\alpha}_1\overline{\alpha}_2  - \overline{\epsilon}_1\overline{\epsilon}_2 + \alpha_4\overline{\alpha}_1 \overline{\epsilon}_2 \right)
 \left( \theta_{12} - \overline{\beta}_1\overline{\beta}_2  - \overline{\lambda}_1\overline{\lambda}_2 + \beta_6\overline{\beta}_1 \overline{\lambda}_2 \right) \nonumber \\
%--------------------
\mathcal{F}_{(3)}^{\fbox{\tiny{11}}} &=&  
 \left[ \theta_{12} - \overline{\alpha}_1\overline{\alpha}_2  - \overline{\kappa}_2\left( \overline{\kappa}_1 -\alpha_5\overline{\alpha}_1 \right)
            - \overline{\delta}_2\left( \overline{\delta}_1 -\alpha_3\overline{\alpha}_1 \right) 
    + \overline{\kappa}_2\delta_5\left( \overline{\delta}_1 -\alpha_3\overline{\alpha}_1 \right) \right]
 \left( \theta_{12} - \overline{\beta}_1\overline{\beta}_2  \right)  \nonumber \\
%--------------------
\mathcal{F}_{(4)}^{\fbox{\tiny{11}}} &=&  
 \left[ \theta_{12} - \overline{\alpha}_1\overline{\alpha}_2  - \overline{\kappa}_2\left( \overline{\kappa}_1 -\alpha_5\overline{\alpha}_1 \right)
            - \overline{\epsilon}_2\left( \overline{\epsilon}_1 -\alpha_4\overline{\alpha}_1 \right) 
    + \overline{\kappa}_2\epsilon_5\left( \overline{\epsilon}_1 -\alpha_4\overline{\alpha}_1 \right) \right]
 \left( \theta_{12} - \overline{\beta}_1\overline{\beta}_2  \right)  \nonumber \\
%--------------------
\mathcal{F}_{(5)}^{\fbox{\tiny{11}}} &=&  
 \left[ \theta_{12} - \overline{\beta}_1\overline{\beta}_2  - \overline{\gamma}_2\left( \overline{\gamma}_1 -\beta_2\overline{\beta}_1 \right)
            - \overline{\chi}_2\left( \overline{\chi}_1 -\beta_7\overline{\beta}_1 \right) 
    + \overline{\gamma}_2\gamma_7\left( \overline{\chi}_1 -\beta_7\overline{\beta}_1 \right) \right]
 \left( \theta_{12} - \overline{\alpha}_1\overline{\alpha}_2  \right) \nonumber \\
%--------------------
\mathcal{F}_{(6)}^{\fbox{\tiny{11}}} &=&  
 \left( \theta_{12} - \overline{\alpha}_1\overline{\alpha}_2  - \overline{\delta}_1\overline{\delta}_2 + \alpha_3\overline{\alpha}_1 \overline{\delta}_2 \right)
 \left( \theta_{12} - \overline{\beta}_1\overline{\beta}_2  - \overline{\chi}_1\overline{\chi}_2 + \beta_7\overline{\beta}_1 \overline{\chi}_2 \right).
\ENA
%ooooooooooooooooooooooooooooooooooooooooooooooooooooooooooooooooooooooooooooooooooooooo

The angular terms for operator $\fbox{12}$ are

\EQA
\mathcal{F}_{(1)}^{\fbox{\tiny{12}}} &=&  
 \left( \theta_{23} - \overline{\alpha}_2\overline{\alpha}_3  - \overline{\beta}_2\overline{\beta}_3 + \alpha_1\overline{\alpha}_2 \overline{\beta}_3 \right)
 \left( \theta_{24} - \overline{\gamma}_2\overline{\gamma}_4  \right) \left( \theta_{34} - \overline{\lambda}_3\overline{\lambda}_4  \right) \nonumber \\
%--------------------
\mathcal{F}_{(2)}^{\fbox{\tiny{12}}} &=&  
 \left( \theta_{34} - \overline{\alpha}_3\overline{\alpha}_4  - \overline{\beta}_3\overline{\beta}_4 + \alpha_1\overline{\alpha}_4 \overline{\beta}_3 \right)
 \left( \theta_{24} - \overline{\epsilon}_2\overline{\epsilon}_4  \right) \left( \theta_{23} - \overline{\lambda}_2\overline{\lambda}_3  \right) \nonumber \\
%--------------------
\mathcal{F}_{(3)}^{\fbox{\tiny{12}}} &=&  
 \left( \theta_{23} - \overline{\alpha}_2\overline{\alpha}_3  - \overline{\beta}_2\overline{\beta}_3 + \alpha_1\overline{\alpha}_3 \overline{\beta}_2 \right)
 \left( \theta_{34} - \overline{\delta}_3\overline{\delta}_4  \right) \left( \theta_{24} - \overline{\kappa}_2\overline{\kappa}_4  \right) \nonumber \\
%--------------------
\mathcal{F}_{(4)}^{\fbox{\tiny{12}}} &=&  
 \left( \theta_{24} - \overline{\alpha}_2\overline{\alpha}_4  - \overline{\beta}_2\overline{\beta}_4 + \alpha_1\overline{\alpha}_4 \overline{\beta}_2 \right)
 \left( \theta_{34} - \overline{\epsilon}_3\overline{\epsilon}_4  \right) \left( \theta_{23} - \overline{\kappa}_2\overline{\kappa}_3  \right) \nonumber \\
%--------------------
\mathcal{F}_{(5)}^{\fbox{\tiny{12}}} &=&  
 \left( \theta_{24} - \overline{\alpha}_2\overline{\alpha}_4  - \overline{\beta}_2\overline{\beta}_4 + \alpha_1\overline{\alpha}_2 \overline{\beta}_4 \right)
 \left( \theta_{23} - \overline{\gamma}_2\overline{\gamma}_3  \right) \left( \theta_{34} - \overline{\chi}_3\overline{\chi}_4  \right) \nonumber \\
%--------------------
\mathcal{F}_{(6)}^{\fbox{\tiny{12}}} &=&  
 \left( \theta_{34} - \overline{\alpha}_3\overline{\alpha}_4  - \overline{\beta}_3\overline{\beta}_4 + \alpha_1\overline{\alpha}_3 \overline{\beta}_4 \right)
 \left( \theta_{23} - \overline{\delta}_2\overline{\delta}_3  \right) \left( \theta_{24} - \overline{\chi}_2\overline{\chi}_4  \right).
\ENA

%ooooooooooooooooooooooooooooooooooooooooooooooooooooooooooooooooooooooooooooooooooooooo

The angular terms for operator $\fbox{13}$ are

\EQA
\mathcal{F}_{(1)}^{\fbox{\tiny{13}}} &=&  
 \left( \theta_{14} - \overline{\alpha}_1\overline{\alpha}_4  - \overline{\gamma}_1\overline{\gamma}_4 + \alpha_2\overline{\alpha}_1 \overline{\gamma}_4 \right)
 \left( \theta_{13} - \overline{\beta}_1\overline{\beta}_3 \right) \left( \theta_{34} - \overline{\lambda}_3\overline{\lambda}_4 \right)  \nonumber \\
%--------------------
\mathcal{F}_{(2)}^{\fbox{\tiny{13}}} &=&  
 \left( \theta_{34} - \overline{\epsilon}_3\overline{\epsilon}_4  - \overline{\lambda}_3\overline{\lambda}_4 + \epsilon_6\overline{\epsilon}_4 \overline{\lambda}_3 \right)
 \left( \theta_{13} - \overline{\beta}_1\overline{\beta}_3 \right) \left( \theta_{14} - \overline{\alpha}_1\overline{\alpha}_4 \right)  \nonumber \\
%--------------------
\mathcal{F}_{(3)}^{\fbox{\tiny{13}}} &=&  
 \left( \theta_{14} - \overline{\beta}_1\overline{\beta}_4  - \overline{\kappa}_1\overline{\kappa}_4 + \beta_5\overline{\beta}_1 \overline{\kappa}_4 \right)
 \left( \theta_{13} - \overline{\alpha}_1\overline{\alpha}_3 \right) \left( \theta_{34} - \overline{\delta}_3\overline{\delta}_4 \right)  \nonumber \\
%--------------------
\mathcal{F}_{(4)}^{\fbox{\tiny{13}}} &=&  
 \left( \theta_{13} - \overline{\beta}_1\overline{\beta}_3  - \overline{\kappa}_1\overline{\kappa}_3 + \beta_5\overline{\beta}_1 \overline{\kappa}_3 \right)
 \left( \theta_{14} - \overline{\alpha}_1\overline{\alpha}_4 \right) \left( \theta_{34} - \overline{\epsilon}_3\overline{\epsilon}_4 \right)  \nonumber \\
%--------------------
\mathcal{F}_{(5)}^{\fbox{\tiny{13}}} &=&  
 \left( \theta_{13} - \overline{\alpha}_1\overline{\alpha}_3  - \overline{\gamma}_1\overline{\gamma}_3 + \alpha_2\overline{\alpha }_1 \overline{\gamma}_3 \right)
 \left( \theta_{14} - \overline{\beta}_1\overline{\beta}_4 \right) \left( \theta_{34} - \overline{\chi}_3\overline{\chi}_4 \right)  \nonumber \\
%--------------------
\mathcal{F}_{(6)}^{\fbox{\tiny{13}}} &=&  
 \left( \theta_{34} - \overline{\delta}_3\overline{\delta}_4  - \overline{\chi}_3\overline{\chi}_4 + \delta_7\overline{\delta}_3 \overline{\chi}_4 \right)
 \left( \theta_{13} - \overline{\alpha}_1\overline{\alpha}_3 \right) \left( \theta_{14} - \overline{\beta}_1\overline{\beta}_4 \right).
\ENA

%ooooooooooooooooooooooooooooooooooooooooooooooooooooooooooooooooooooooooooooooooooooooo

The angular terms for operator $\fbox{14}$ are

\EQA
\mathcal{F}_{(1)}^{\fbox{\tiny{14}}} &=&  
 \left( \theta_{14} - \overline{\beta}_1\overline{\beta}_4 - \overline{\lambda}_1\overline{\lambda}_4 + \beta_6\overline{\beta}_1 \overline{\lambda}_4 \right) 
 \left( \theta_{12} - \overline{\alpha}_1\overline{\alpha}_2  \right) \left( \theta_{24} - \overline{\gamma}_2\overline{\gamma}_4  \right) \nonumber \\
%--------------------
\mathcal{F}_{(2)}^{\fbox{\tiny{14}}} &=&  
 \left( \theta_{12} - \overline{\beta}_1\overline{\beta}_2  - \overline{\lambda}_1\overline{\lambda}_2 + \beta_6\overline{\beta}_1 \overline{\lambda}_2 \right) 
 \left( \theta_{14} - \overline{\alpha}_1\overline{\alpha}_4 \right) \left( \theta_{24} - \overline{\epsilon}_2\overline{\epsilon}_4  \right) \nonumber \\
%--------------------
\mathcal{F}_{(3)}^{\fbox{\tiny{14}}} &=&  
 \left( \theta_{14} - \overline{\alpha}_1\overline{\alpha}_4 - \overline{\delta}_1\overline{\delta}_4 + \alpha_3\overline{\alpha}_1 \overline{\delta}_4  \right)
 \left( \theta_{12} - \overline{\beta}_1\overline{\beta}_2  \right) \left( \theta_{24} - \overline{\kappa}_2\overline{\kappa}_4  \right) \nonumber \\
%--------------------
\mathcal{F}_{(4)}^{\fbox{\tiny{14}}} &=&  
 \left( \theta_{24} - \overline{\epsilon}_2\overline{\epsilon}_4  - \overline{\kappa}_2\overline{\kappa}_4 + \epsilon_5\overline{\epsilon}_4 \overline{\kappa}_2 \right)
 \left( \theta_{12} - \overline{\beta}_1\overline{\beta}_2  \right) \left( \theta_{14} - \overline{\alpha}_1\overline{\alpha}_4  \right) \nonumber \\
%--------------------
\mathcal{F}_{(5)}^{\fbox{\tiny{14}}} &=&  
 \left( \theta_{24} - \overline{\gamma}_2\overline{\gamma}_4 - \overline{\chi}_2\overline{\chi}_4 + \gamma_7\overline{\gamma}_2\overline{\chi}_4 \right)
 \left( \theta_{12} - \overline{\alpha}_1\overline{\alpha}_2  \right) \left( \theta_{14} - \overline{\beta}_1\overline{\beta}_4  \right) \nonumber \\
%--------------------
\mathcal{F}_{(6)}^{\fbox{\tiny{14}}} &=&  
 \left( \theta_{12} - \overline{\alpha}_1\overline{\alpha}_2  - \overline{\delta}_1\overline{\delta}_2 + \alpha _3\overline{\alpha}_1 \overline{\delta}_2 \right) 
 \left( \theta_{14} - \overline{\beta}_1\overline{\beta}_4 \right) \left( \theta_{24} - \overline{\chi}_2\overline{\chi}_4  \right). 
\ENA

%ooooooooooooooooooooooooooooooooooooooooooooooooooooooooooooooooooooooooooooooooooooooo

The angular terms for operator $\fbox{15}$ are

\EQA
\mathcal{F}_{(1)}^{\fbox{\tiny{15}}} &=&  
 \left( \theta_{23} - \overline{\gamma}_2\overline{\gamma}_3 - \overline{\lambda}_2\overline{\lambda}_3 + \gamma_6\overline{\gamma}_2\overline{\lambda}_4 \right)
 \left( \theta_{12} - \overline{\alpha}_1\overline{\alpha}_2  \right) \left( \theta_{13} - \overline{\beta}_1\overline{\beta}_3  \right) \nonumber \\
%--------------------
\mathcal{F}_{(2)}^{\fbox{\tiny{15}}} &=&  
 \left( \theta_{12} - \overline{\alpha}_1\overline{\alpha}_2  - \overline{\epsilon}_1\overline{\epsilon}_2 + \alpha_4\overline{\alpha}_1 \overline{\epsilon}_2 \right)
 \left( \theta_{13} - \overline{\beta}_1\overline{\beta}_3 \right) \left( \theta_{23} - \overline{\lambda}_2\overline{\lambda}_3  \right) \nonumber \\
%--------------------
\mathcal{F}_{(3)}^{\fbox{\tiny{15}}} &=&  
 \left( \theta_{23} - \overline{\delta}_2\overline{\delta}_3 - \overline{\kappa}_2\overline{\kappa}_3 + \delta_5\overline{\delta}_3\overline{\kappa}_2 \right)
 \left( \theta_{12} - \overline{\beta}_1\overline{\beta}_2  \right) \left( \theta_{13} - \overline{\alpha}_1\overline{\alpha}_3  \right) \nonumber \\
%--------------------
\mathcal{F}_{(4)}^{\fbox{\tiny{15}}} &=&  
 \left( \theta_{13} - \overline{\alpha}_1\overline{\alpha}_3  - \overline{\epsilon}_1\overline{\epsilon}_3 + \alpha_4\overline{\alpha}_1 \overline{\epsilon}_3 \right)
 \left( \theta_{12} - \overline{\beta}_1\overline{\beta}_2 \right) \left( \theta_{23} - \overline{\kappa}_2\overline{\kappa}_3 \right) \nonumber \\
%--------------------
\mathcal{F}_{(5)}^{\fbox{\tiny{15}}} &=&  
 \left( \theta_{13} - \overline{\beta}_1\overline{\beta}_3  - \overline{\chi}_1\overline{\chi}_3 + \beta_7\overline{\beta}_1 \overline{\chi}_3 \right)
 \left( \theta_{12} - \overline{\alpha}_1\overline{\alpha}_2 \right) \left( \theta_{23} - \overline{\gamma}_2\overline{\gamma}_3 \right)  \nonumber \\
%--------------------
\mathcal{F}_{(6)}^{\fbox{\tiny{15}}} &=&  
 \left( \theta_{12} - \overline{\beta}_1\overline{\beta}_2  - \overline{\chi}_1\overline{\chi}_2 + \beta_7\overline{\beta}_1 \overline{\chi}_2 \right) 
 \left( \theta_{13} - \overline{\alpha}_1\overline{\alpha}_3 \right) \left( \theta_{23} - \overline{\delta}_2\overline{\delta}_3  \right). 
\ENA

%ooooooooooooooooooooooooooooooooooooooooooooooooooooooooooooooooooooooooooooooooooooooo

Finally, the angular terms for operator $\fbox{16}$ are

\EQA
 \mathcal{F}_{(1)}^{\fbox{\tiny{16}}}&=&  \left( \theta_{12} -\overline{\alpha}_1 \overline{\alpha}_2 \right) 
                                            \left( \theta_{13} -\overline{\beta}_1 \overline{\beta}_3 \right) 
                                             \left( \theta_{24} -\overline{\gamma}_2 \overline{\gamma}_4 \right) 
                                              \left( \theta_{34} -\overline{\lambda}_3 \overline{\lambda}_4 \right) \nonumber \\
 \mathcal{F}_{(2)}^{\fbox{\tiny{16}}}&=&  \left( \theta_{14} -\overline{\alpha}_1 \overline{\alpha}_4 \right)
                                            \left( \theta_{13} -\overline{\beta}_1 \overline{\beta}_3 \right)
                                             \left( \theta_{24} -\overline{\epsilon}_2 \overline{\epsilon}_4 \right) 
                                              \left( \theta_{23} -\overline{\lambda}_2 \overline{\lambda}_3 \right)  \nonumber \\
 \mathcal{F}_{(3)}^{\fbox{\tiny{16}}}&=&  \left( \theta_{13} -\overline{\alpha}_1 \overline{\alpha}_3 \right)
                                            \left( \theta_{12} -\overline{\beta}_1 \overline{\beta}_2 \right)
                                             \left( \theta_{34} -\overline{\delta}_3 \overline{\delta}_4 \right) 
                                              \left( \theta_{24} -\overline{\kappa}_2 \overline{\kappa}_4 \right)  \nonumber \\
 \mathcal{F}_{(4)}^{\fbox{\tiny{16}}}&=&  \left( \theta_{14} -\overline{\alpha}_1 \overline{\alpha}_4 \right)
                                            \left( \theta_{12} -\overline{\beta}_1 \overline{\beta}_2 \right)
                                             \left( \theta_{34} -\overline{\epsilon}_3 \overline{\epsilon}_4 \right) 
                                              \left( \theta_{23} -\overline{\kappa}_2 \overline{\kappa}_3 \right)  \nonumber \\
 \mathcal{F}_{(5)}^{\fbox{\tiny{16}}}&=&  \left( \theta_{12} -\overline{\alpha}_1 \overline{\alpha}_2 \right)
                                            \left( \theta_{14} -\overline{\beta}_1 \overline{\beta}_4 \right)
                                             \left( \theta_{23} -\overline{\gamma}_2 \overline{\gamma}_3 \right) 
                                              \left( \theta_{34} -\overline{\chi}_3 \overline{\chi}_4 \right)  \nonumber \\
 \mathcal{F}_{(6)}^{\fbox{\tiny{16}}}&=&  \left( \theta_{13} -\overline{\alpha}_1 \overline{\alpha}_3 \right)
                                            \left( \theta_{14} -\overline{\beta}_1 \overline{\beta}_4 \right)
                                             \left( \theta_{23} -\overline{\delta}_2 \overline{\delta}_3 \right) 
                                              \left( \theta_{24} -\overline{\chi}_2 \overline{\chi}_4 \right).   
% \label{ang_terms_Pi_16}
\ENA

\end{widetext}

%LLLLLLLLLLLLLLLLLLLLLLLLLLLLLLLLLLLLLLLLLLLLLLLLLLLLLLLLLLLLLLLLLLLLLLLLLLLLLLLLLLLLLLLLLLLLL
%LLLLLLLLLLLLLLLLLLLLLLLLLLLLLLLLLLLLLLLLLLLLLLLLLLLLLLLLLLLLLLLLLLLLLLLLLLLLLLLLLLLLLLLLLLLLL

\end{document}